\documentclass[10pt]{article}

\usepackage[T1]{fontenc}
\usepackage{lmodern}
\usepackage[a4paper, margin=2.5cm]{geometry}
\usepackage{graphicx}
\usepackage{amsmath}
\usepackage{amssymb}
\usepackage{hyperref}
\usepackage[numbers,sort&compress]{natbib}
\usepackage{array}
\usepackage{longtable}
\usepackage{booktabs}
\usepackage{pdflscape}
\usepackage{multirow}
\usepackage{xcolor}
\definecolor{paperred}{HTML}{B3264F}
\definecolor{paperblue}{HTML}{1B6CA8}
\usepackage{caption}
\usepackage{float}
\usepackage[utf8]{inputenc}
\usepackage{orcidlink}
\usepackage{tikz}
\usetikzlibrary{arrows.meta,positioning,shapes.geometric,decorations.pathmorphing,decorations.pathreplacing}
\providecommand{\ket}[1]{|#1\rangle}

\hypersetup{
  pdftitle={QOuLiPo: What a quantum computer sees when it reads a book},
  pdfauthor={Christophe Jurczak},
  hidelinks,
}

\begin{document}

% ─── Title block ───────────────────────────────────────────────────────────

\begin{center}
  {\Large\bfseries QOuLiPo: What a quantum computer sees when it reads a book}

  \vspace{0.6em}

  \vspace{1em}

  {\large Christophe Jurczak\footnote{\href{mailto:christophe@quantonation.com}{christophe@quantonation.com}} \orcidlink{0009-0001-8204-7333}}

  \vspace{0.4em}

  {\footnotesize Quantonation, Dallas, TX, USA}

  \vspace{0.4em}

  {\footnotesize\itshape \today}

  \vspace{1em}

  \begin{minipage}{0.85\textwidth}
    {\small
      \textbf{Abstract.} What does a book look like to a quantum computer? This paper takes eight classical works of the Renaissance and its late-antique inheritance --- from Augustine through Dante, Giambullari, and Galileo --- and runs each one through a neutral-atom quantum processor. The bridge is the mathematics of graphs. Encoding texts as graphs is by now standard in the digital humanities; what is new is that the graph is not a representation the computer manipulates but the physical configuration the hardware itself realises. Each textual unit (page, canto, chapter, paragraph, depending on the work) becomes an atom; for engineered exact unit-disk designs, graph edges \emph{are} physical blockade constraints, while for natural texts the processor receives the best 2D unit-disk approximation to the semantic graph. A question literary scholars actually ask is then read directly off the atoms: which subset of a book's units is mutually so distinct that no additional unit can be added without violating their mutual independence?

      Three contributions follow. First, we introduce \emph{rigidity} $\rho$, a new metric that measures how unique a book's structural backbone is --- discriminating Marguerite de Navarre's \emph{Heptam\'eron} (rigid, twelve-nouvelle hard core) from Boethius's \emph{De Consolatione Philosophiae} (fully fungible, every chapter substitutable). Second, we invert the pipeline: rather than extracting a graph from existing prose, we pick a target graph the hardware encodes natively, and \emph{write} a book whose structure matches it. The twenty-nine texts written this way, collected under the name \emph{QOuLiPo}, extend the OuLiPo tradition to graph-topological constraints and, together with the eight natural texts of the corpus, form a benchmark distribution against which neutral-atom hardware can be tracked as it scales. Third, we run both natural and engineered texts on Pasqal's FRESNEL processor at sizes up to one hundred atoms: the hardware behaves faithfully across the corpus and, on engineered texts purpose-built to match its native geometry, reaches high approximation ratios and, on the cleanest instances, returns the exact backbone; on natural texts the semantic graph must be embedded approximately into the planar register, exposing that embedding step as the present bottleneck and better embedders as the path to fuller recovery. The framework already opens to a new mode the hardware natively supports --- three-dimensional registers that escape the planar embedding bottleneck.

      A cloud-accessible quantum machine plus an agentic coding environment now lets a single investigator run this pipeline end-to-end. What is reported is an application layer, not a speedup --- humanistic instances ready to load onto neutral-atom processors as they scale, already complementing classical text analysis. The Digital Humanities community has a stake in building familiarity with this hardware now: the engineered-corpus design choices made today fix the benchmark distribution future hardware will be measured against.
    }
  \end{minipage}
\end{center}

\vspace{1em}
\noindent\rule{\textwidth}{0.5pt}
\vspace{1em}

% ═══════════════════════════════════════════════════════════════════════════
%  MAIN PAPER
% ═══════════════════════════════════════════════════════════════════════════

% ─── Introduction ────────────────────────────────────────────────────────

\section{Introduction}\label{sec:intro}

The Maximum Independent Set (MIS) problem asks for the largest subset of mutually non-adjacent vertices in a graph. In physics, MIS occupies a privileged position on \emph{unit-disk graphs}: graphs whose vertices can be drawn as points in the plane such that two vertices are connected if and only if they lie within a fixed distance. An array of cold atoms (typically rubidium, caesium, or ytterbium) held in optical tweezers, where no two atoms within a blockade radius can simultaneously occupy an excited state, can be driven by an adiabatic ramp of the laser detuning into a configuration that solves MIS on the unit-disk graph defined by the array, as its lowest-energy arrangement under the Rydberg Hamiltonian. No software compilation sits between the problem and the hardware. The physical setting is the \emph{Rydberg Hamiltonian} on a neutral-atom register \cite{henriet2020quantum,browaeys2020many}; quantum MIS on this platform has been proposed \cite{pichler2018quantum} and demonstrated \cite{ebadi2022quantum,cazals2025mis}. The hardware that realises it is the neutral-atom quantum processors built by several companies.

\begin{figure}[!htbp]
\centering
\begin{tikzpicture}[
    >=Stealth, font=\small,
    atom/.style={circle, draw=black!70, fill=black!10, inner sep=1.6pt},
    atomr/.style={circle, draw=paperred, fill=paperred, inner sep=1.8pt},
    level/.style={thick},
    laser/.style={decorate, decoration={snake, amplitude=0.6pt, segment length=3pt}, ->, paperred},
    edgeln/.style={gray!70, thin},
  ]
  %========== Panel (a): atomic levels + blockade ==========
  \begin{scope}[shift={(0,0)}]
    % (a) label drawn outside the scope, see end of tikzpicture
    % two atoms
    \shade[ball color=paperred!20, opacity=0.5] (0,1.5) circle (1.45);
    \draw[paperred, dashed] (0,1.5) circle (1.45);
    \node[paperred,font=\scriptsize] at (1.05,2.4) {$R_b$};
    \node[atomr, label={[paperred,font=\scriptsize]above:$\ket{r}$}] (ai) at (-0.65,1.5) {};
    \node[atom,  label={[font=\scriptsize]above:$\ket{g}$}] (aj) at ( 0.65,1.5) {};
    \node[font=\scriptsize] at (-0.65,1.05) {$i$};
    \node[font=\scriptsize] at ( 0.65,1.05) {$j$};
    \draw[<->, thin] (-0.65,0.65) -- node[below,font=\scriptsize]{$r_{ij}<R_b$} (0.65,0.65);
    % energy ladder below
    \begin{scope}[shift={(0,-2.0)}]
      \draw[level] (-1.25,0) -- (-0.75,0) node[right,font=\scriptsize]{$\ket{gg}$};
      \draw[level] (-1.25,0.7) -- (-0.75,0.7) node[right,font=\scriptsize]{$(\ket{gr}{+}\ket{rg})/\sqrt{2}$};
      \draw[level,paperred,very thick] (-1.25,1.9) -- (-0.75,1.9)
          node[right,font=\scriptsize,black]{$\ket{rr}$ shifted by $U{=}C_6/r_{ij}^6$};
      \draw[laser] (-1.00,0.05) -- (-1.00,0.65);
      \node[paperred,font=\scriptsize] at (-1.50,0.35) {$\Omega$};
      \draw[<->, thin, gray] (-0.55,0.75) -- (-0.55,1.85);
      \node[gray,font=\scriptsize] at (-0.25,1.3) {$\gg \Omega$};
    \end{scope}
  \end{scope}
  %========== Panel (b): tweezer array (FRESNEL-style triangular lattice) + atoms ==========
  \begin{scope}[shift={(7.6,0)}, scale=1.35, transform shape]
    % (b) label drawn outside the scope, see end of tikzpicture
    % --- Triangular trap lattice (rhombic/row-offset) -------------------
    % Lattice constant a = 1.1 (in scope units); row spacing = a*sqrt(3)/2 ≈ 0.95.
    % Two interleaved row sets at y ∈ {2.6, 0.7} (offset 0)
    \foreach \xi in {-2.2,-1.1,0,1.1,2.2}{
      \foreach \yi in {2.6, 0.7}{
        \node[circle, draw=black!28, fill=none, inner sep=0pt, minimum size=3.4pt] at (\xi,\yi) {};
      }
    }
    % and y ∈ {1.65, -0.25} (offset a/2 ≈ 0.55)
    \foreach \xi in {-1.65,-0.55,0.55,1.65}{
      \foreach \yi in {1.65, -0.25}{
        \node[circle, draw=black!28, fill=none, inner sep=0pt, minimum size=3.4pt] at (\xi,\yi) {};
      }
    }
    % --- Atoms loaded at 9 of the trap sites: connected graph, MIS = 4 ---
    \coordinate (a1) at (-1.1,  2.6);   % red, MIS
    \coordinate (a2) at (-0.55, 1.65);  % ground
    \coordinate (a3) at ( 0.55, 1.65);  % red, MIS
    \coordinate (a4) at ( 1.1,  2.6);   % ground
    \coordinate (a5) at ( 2.2,  2.6);   % red, MIS
    \coordinate (a6) at ( 1.65, 1.65);  % ground
    \coordinate (a7) at (-0.55,-0.25);  % red, MIS
    \coordinate (a8) at ( 0.55,-0.25);  % ground
    \coordinate (a9) at ( 0.0,  0.7);   % ground
    % --- Blockade edges: every loaded pair at NN distance (a) -----------
    \draw[edgeln] (a1)--(a2);
    \draw[edgeln] (a2)--(a3);
    \draw[edgeln] (a3)--(a4);
    \draw[edgeln] (a4)--(a5);
    \draw[edgeln] (a3)--(a6);
    \draw[edgeln] (a4)--(a6);
    \draw[edgeln] (a5)--(a6);
    \draw[edgeln] (a2)--(a9);
    \draw[edgeln] (a3)--(a9);
    \draw[edgeln] (a7)--(a9);
    \draw[edgeln] (a8)--(a9);
    \draw[edgeln] (a7)--(a8);
    % --- One R_b indicator (illustrative) on a MIS atom ----------------
    \draw[paperred!55, dashed, line width=0.4pt] (a3) circle (1.15);
    \draw[<-, paperred, line width=0.3pt] (a3) -- ++(1.15,0)
          node[right, font=\scriptsize, paperred] {$R_b$};
    % --- Atoms drawn on top of the trap markers -------------------------
    \node[atomr] at (a1) {};
    \node[atom]  at (a2) {};
    \node[atomr] at (a3) {};
    \node[atom]  at (a4) {};
    \node[atomr] at (a5) {};
    \node[atom]  at (a6) {};
    \node[atomr] at (a7) {};
    \node[atom]  at (a8) {};
    \node[atom]  at (a9) {};
    % --- Legend ---------------------------------------------------------
    \node[font=\scriptsize,align=center] at (0,-1.55)
      {\textcolor{paperred}{$\bullet$} $\ket{r}$ atom in MIS\quad
       \textcolor{black!70}{$\circ$} $\ket{g}$ ground-state atom\quad
       {\tikz[baseline=-0.4ex]{\node[circle,draw=black!28,fill=none,inner sep=0pt,minimum size=3.4pt]{};}} unloaded trap};
  \end{scope}
  % ── panel labels at uniform size, on the left of each subpicture ──
  \node[font=\bfseries] at (-2.4, 3.0) {(a)};
  \node[font=\bfseries] at ( 4.5, 3.0) {(b)};
\end{tikzpicture}
\caption{Neutral-atom MIS in one picture. (a) \emph{Rydberg blockade.} Two atoms held within a blockade radius $R_b$ cannot both be driven to the excited Rydberg state $\ket{r}$: the doubly-excited level $\ket{rr}$ is shifted up by an interaction energy $U = C_6/r_{ij}^6$ much larger than the Rabi coupling $\Omega$ that the laser would need to populate it, so only the symmetric singly-excited combination $(\ket{gr}+\ket{rg})/\sqrt{2}$ is reachable. The blockade radius $R_b$ is the distance at which $U(R_b) = \hbar\Omega$; for typical $\Omega/2\pi \sim 1$--$5$~MHz on rubidium-87, $R_b \approx 5$--$10\,\mu$m. (b) \emph{Atom register = text graph.} The faint open circles are an optical-tweezer lattice (FRESNEL uses a triangular trap layout); atoms are loaded into a chosen subset, and that loaded subset is the register that runs the computation. Each loaded atom represents one textual unit (a page, a canto, a chapter); the gray edges connect every loaded pair within blockade range $R_b$ and are therefore the edges of the underlying graph. The register's ground state under the Rydberg drive is a Maximum Independent Set of that graph: a largest subset of atoms (red, in $\ket{r}$) with no two within $R_b$ of each other. The dashed circles around the red atoms mark their blockade exclusion zones --- no second atom can be excited inside them, which is exactly the independence constraint. The register is not a representation of the graph: it \emph{is} the graph, encoded in matter. After \cite{ebadi2022quantum,browaeys2020many}.}
\label{fig:rydberg_mis}
\end{figure}
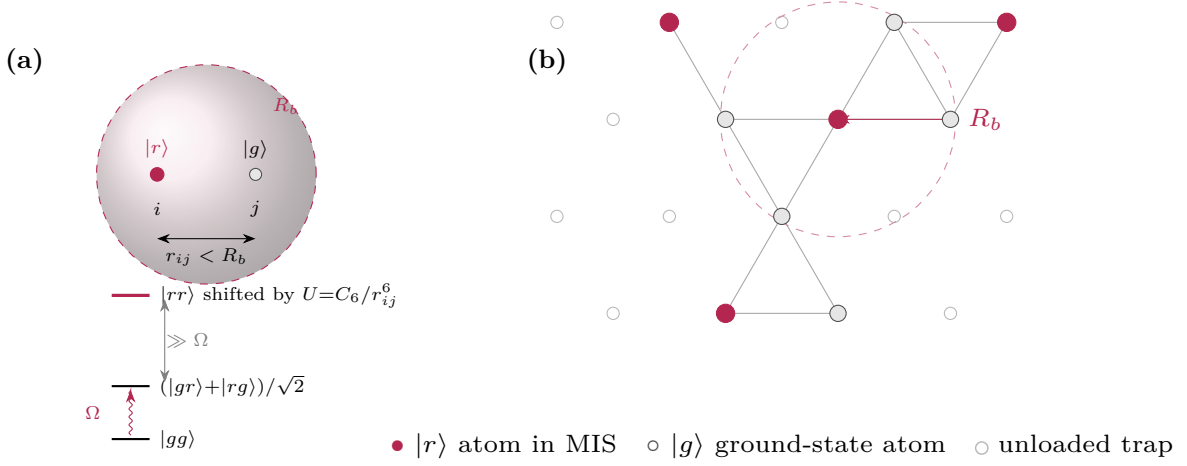

This mapping is not a metaphor. The Rydberg blockade \emph{is} the independence constraint, enforced by the physics: any two atoms within $R_b$ cannot both be driven to the Rydberg state, exactly as the Independent Set constraint forbids any two adjacent vertices from both being selected. Laser detuning plays the role of a per-vertex inclusion reward, and the ground state prepared by the adiabatic protocol is the MIS of the blockade-adjacency graph \cite{henriet2020quantum,browaeys2020many,jaksch2000rydberg,urban2009observation}. The hardware, the register geometry, and a concrete MIS read-out are shown in Figure~\ref{fig:rydberg_mis}.

Here we bring MIS to an unexpected domain: the structural analysis of literary texts, specifically the books that the Renaissance recovered and printed. Any book can be decomposed into units (nouvelles, cantos, chapters, printed pages or paragraphs), each embedded as a vector in a high-dimensional semantic space. Connecting units whose embeddings are sufficiently similar produces a graph, and the MIS of that graph selects what we call the \emph{structural backbone} of the book: the largest set of pairwise-dissimilar units, with every omitted unit adjacent to at least one selected one.

A graph can have many distinct optimal MIS solutions, all of the same size. We introduce \emph{rigidity} $\rho \in [0, 1]$ as the fraction of backbone nodes common to every optimal solution: $\rho = 1$ means the backbone is unique (every selected unit is irreplaceable); $\rho = 0$ means the backbone is maximally degenerate (any element can be substituted by another structurally equivalent one). Rigidity is the central new quantity of this paper: it characterises the space of possible structures a text permits.

\paragraph{Contributions.}
\emph{(i)} We establish MIS as a structural analysis tool for historical texts in their original languages \emph{under a fixed multilingual sentence-transformer embedding} (\texttt{multilingual-\allowbreak e5-\allowbreak large-\allowbreak instruct}), validating it across eight corpora spanning four centuries and three languages, and showing that rigidity $\rho$, a property of the (text, embedding) pair, discriminates between qualitatively different kinds of textual organisation. The results are striking: Dante's \emph{Inferno} is moderately rigid ($\rho = 0.444$); Marguerite de Navarre's \emph{Heptam\'eron} at $k = 16$ has a 12-nouvelle hard core in every optimum ($\rho = 0.857$); Boethius's \emph{De Consolatione} is fully fungible ($\rho = 0$, hundreds of optima whose intersection is empty). Backbone \emph{size} ($\mathrm{MIS}/N$) varies moderately (0.10--0.30); rigidity varies dramatically and reveals structural properties invisible to word-frequency methods \cite{rockwell2016hermeneutica} and topic models \cite{blei2003lda}.
\emph{(ii)} We invert the pipeline: rather than extract a graph from given prose, we pick a target graph with prescribed properties (unique MIS by double-domination, $3^{17}$ equivalent backbones by disjoint cliques, exact unit-disk realisations on king lattices, a dodecahedron by 30-motif edge-assignment, and more) and \emph{write} prose so the page-level adjacencies match the target. This extends OuLiPo (the \emph{Ouvroir de Litt\'erature Potentielle}, founded 1960) \cite{queneau1961sonnets,perec1969disparition,perec1978vie,berkman2022oulipo} to graph-topological constraints; we engineer twenty-nine such texts at register sizes (16--100 atoms) matched to the hardware. The designed graphs are ground truth by construction, so the engineered corpus is also a calibration instrument for the embedder, the classical Integer Linear Programming (ILP) solver, and the Quantum Processing Unit (QPU) pipeline.
\emph{(iii)} We run both natural and engineered texts on Pasqal's FRESNEL\footnote{The FRESNEL processor is named after Augustin-Jean Fresnel (1788--1827), the French physicist whose wave theory of light, Fresnel equations, and stepped-lens design are foundational to optics and to the laser-based atom-trapping techniques the machine relies on.} neutral-atom QPU across register sizes $N = 25$--$100$. On every row the approximation ratio against the canonical adjacency the QPU solves sits in the published $0.84$--$1.00$ band of neutral-atom MIS benchmarks \cite{ebadi2022quantum,cazals2025mis} --- exact-UDG engineered rows and simulated-annealing (SA) embedded $k$-NN rows alike. The recovery rate of the underlying \emph{text}-MIS is bounded above by SA recall on the natural-text rows: the upstream 2D embedding step is what loses information, not the processor.

The Renaissance was itself a technology revolution applied to texts: humanist editors collated manuscripts, the Aldine press stabilised them in pocket octavos, and a body of classical and contemporary works that had been scattered across monastic libraries became reproducible, comparable, and portable. Applying today's most advanced computational technology, neutral-atom quantum processors, to those same objects is therefore not an arbitrary pairing but a continuation: each generation reads the inherited corpus through the most analytic tools it has. We read these same texts (Ausonius, Dante, Marguerite de Navarre, Giambullari, Augustine, Lactantius, Boethius, Galileo) and ask which admit a unique structural backbone, which admit many, and whether a quantum processor can recover that backbone by physical means.

\paragraph{Methodology: a single-investigator agentic loop.} Three developments now make such bridging possible at the scale of a single investigator. Cloud access exposes the machines; large-language-model (LLM) agents now translate a domain question (``what are the structural chapters of this book?'') through the full pipeline (text $\to$ embedding $\to$ $k$-NN graph $\to$ atom register $\to$ laser pulse), submit it, and explain the result back in the vocabulary of the originating field. Shiraishi \emph{et al.}\ \cite{shiraishi2026mcp} demonstrated an LLM-agent interface to quantum emulators built on the Model Context Protocol (MCP), Anthropic's open standard for connecting language-model agents to external tools and services; this paper is the Pasqal/FRESNEL instantiation. Concretely, every QPU and EMU\_MPS (Pasqal Cloud's matrix-product-state classical emulator of the Rydberg Hamiltonian) submission reported here was authored through an agentic loop with Anthropic's Claude Opus (versions 4.6 and 4.7, 1-million-token context for the larger campaigns): the agent reads the text-graph artefacts, calls Pulser (Pasqal's open-source Python library for designing atom registers and pulse sequences \cite{silverio2022pulser}) to construct the register and pulse, submits to the Pasqal Cloud, and reports approximation ratio, validity, and Hamming-weight statistics. The loop runs end-to-end through standard Bash, Python, and file-system tool calls; an MCP-style native interface to the Pasqal Cloud, analogous to \cite{shiraishi2026mcp}, would streamline this further but is not a prerequisite. Operationally, the cloud QPU was stable enough to support an iterative single-investigator workflow on a fixed timeline: FRESNEL\_CAN1, hosted at the DistriQ quantum innovation zone in Sherbrooke, Qu\'ebec, was available from the user's perspective every day except for planned maintenance, with no submission lost to an unscheduled outage, and no result in this paper depends on privileged hardware access. Day-to-day reliability of this kind is the precondition without which a single-investigator agentic loop is not viable in practice; it is what lets a project plan around quantum hardware in the same way it would plan around any other piece of cloud infrastructure. Taken together, these three pieces (agentic interface, mature SDK, reliable cloud hardware) bring the barrier to entry to quantum hardware lower than commonly assumed: with agentic LLM tooling on top of a stable cloud QPU, domain specialists (literary scholars, historians, philologists) can drive a real QPU end-to-end without first becoming quantum physicists.

\paragraph{What this paper is, and is not.} This paper is deliberately written as a boundary object between quantum computing and computational literary studies \cite{moretti2013distant,jockers2013macroanalysis,underwood2019distant}. Its contribution is not a new quantum algorithm nor a finished literary reading. It is an application-layer and corpus proposal: a reproducible interface between text structure and hardware-native neutral-atom MIS, plus a benchmark distribution (the QOuLiPo corpus) calibrated against the published $(N, d)$ neutral-atom MIS landscape of Cazals \emph{et al.}\ \cite{cazals2025mis}, where $N$ is register size and $d$ graph density. The pieces a humanist reader needs (a deposited corpus, an ILP-verified backbone, a falsifiable rigidity metric) and the pieces a physics reader needs (an exact-UDG benchmark family, a hardware-native pipeline, a regime where classical simulation hits its ceiling) sit on the same object. Static 2D-UDG MIS admits a polynomial-time approximation scheme (PTAS) and is not a quantum-utility target at our sizes (MIS on general graphs remains NP-hard); what this paper establishes, on the way to the harder regimes addressed in \S\ref{sec:beyond2d}, is that the hardware behaves faithfully on the corpus. The paper should be read as a manifesto for a category that did not exist before, populated densely enough to be tested.

\medskip

This contribution sits at the intersection of two strands of prior work. On the text side, MIS has been applied to extractive text summarisation before \cite{uckan2020extractive,hark2025mis}, but not as a structural-analysis tool for literary corpora, and never as far as we're aware on quantum hardware. Graph-based representations of texts have a growing track record in computational linguistics: complex-network features correlate with text quality and authorship \cite{antiqueira2009complex,amancio2013network,amancio2015probing}; character-interaction networks reveal narrative structure \cite{elson2010extracting,moretti2011network,labatut2019extraction}; grammar-aware quantum natural-language-processing (NLP) models have been run on gate-based hardware for sentence classification \cite{meichanetzidis2023grammar}. Our approach differs: unit-level semantic similarity builds the graph, and MIS extracts a whole-graph backbone rather than ranking individual nodes.

A handful of bridges between quantum computing and the humanities have already been built. Barzen and Leymann \cite{barzen2019quantum,barzen2021digital} have called for a ``quantum humanities'' research programme; Miranda's edited volume \cite{miranda2022qc_arts_humanities} collects the existing work. Published hardware experiments concentrate in two small clusters: quantum natural-language processing (QNLP) on superconducting and trapped-ion processors, scaling from toy sentence classification to hundred-sentence experiments \cite{meichanetzidis2023grammar,lorenz2023qnlp,wazni2024pronoun}, and quantum-assisted music composition collected in Miranda's volume. QOuLiPo is distinct from the QNLP lineage in its compute modality (analog Rydberg-blockade MIS on an atom register rather than parameterised quantum circuits) and extends the platform base of the field by adding neutral atoms to the hardware already represented. The pipeline is end-to-end on real hardware, with constrained writing as the inverse output of a designed graph for the engineered half of the corpus.

% ─── Pipeline ──────────────────────────────────────────────────────────────

\section{From Text to Graph}\label{sec:pipeline}

\begin{figure}[!b]
\centering
\includegraphics[width=\textwidth]{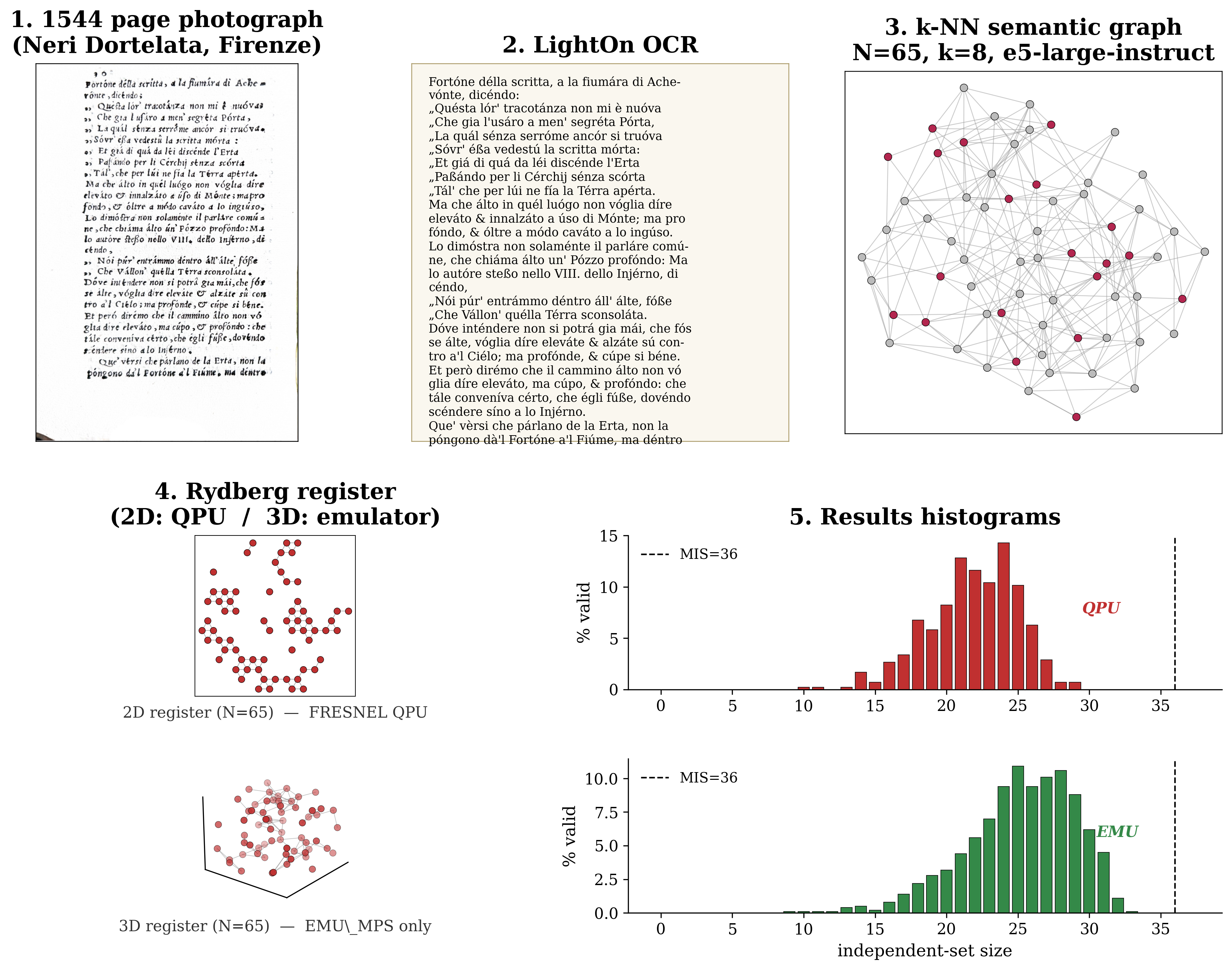}
\caption{End-to-end pipeline on Giambullari's 1544 treatise: (1) original page photograph, (2) fine-tuned vision-language-model OCR (VLM-OCR) at 2.48\% character error rate (CER) \cite{jurczak2026lightonocr}, (3) $k$-NN text graph, (4) 2D QPU and 3D emulator registers, (5) bitstring histograms.}
\label{fig:giambullari_workflow}
\end{figure}

Figure~\ref{fig:giambullari_workflow} shows the end-to-end pipeline on a 1544 book, from page photograph to bitstring histogram. The graph-construction step itself (which is what this section describes) proceeds through four stages. First, the source text is segmented into $N$ units. We adopt a tiered granularity rule: when the author divided the text into discrete structural units (cantos, nouvelles, prosimetric pieces, poems, chapters, paragraphs --- per text, see Appendix~\ref{app:corpus}), we use those. When no such division exists and page-faithful alignment with the original edition is available, we fall back to the printed page --- the smallest division the historical reader actually encountered. For long continuous prose whose authorial division is too coarse to give the pipeline statistical resolution (Galileo's \emph{Dialogo}, four days of dialogue at $N{=}4$, being the canonical case), we fall back to fixed analytic windows --- $400$ words for Galileo, chosen as the median word count of a printed page of the 1632 Landini \emph{editio princeps}, so that the analytic window remains commensurate with what the historical reader encountered. The choice of unit per text is reported explicitly in Appendix~\ref{app:corpus}.

Sourcing faithful transcriptions of early-modern texts in their original language is harder than it sounds. Two failure modes appear. First, the text exists online but only as a modernised critical edition: spelling normalised, long-s resolved, ligatures expanded. Ausonius is representative: his poetry is freely available through the Loeb Classical Library \cite{loeb}, The Latin Library \cite{latinlibrary}, and DigiLiBLT \cite{digiliblt}, but no open transcription preserves the orthography of the Aldine 1517 \emph{editio princeps} we actually photographed. Second, no open-access transcription exists at all. Giambullari's 1544 \emph{De'l sito, forma, \& misure dello Inferno di Dante} (Florence, Neri Dortelata) is the Accademia Fiorentina's first systematic reconstruction of Dante's infernal topography --- building on Antonio Manetti's earlier geometric attempt (c.~1480s, printed posthumously by Giunti in 1506) and Landino's 1481 commentary, and feeding directly into the lineage that culminates in Galileo's 1587 Florentine Academy lectures on the geometry of Dante's Inferno \cite{levyleblond2022dante,peterson2011galileo} --- yet no digital edition of the \emph{editio princeps} is publicly available: not in Bibliotheca Italiana, not in the Accademia della Crusca, not on Wikisource, not on Google Books.

Both gaps are closed by a single per-edition low-rank-adaptation (LoRA) fine-tune of a vision-language model on a handful of manually-corrected reference pages, run on photographs taken with a consumer smartphone rather than a flatbed scanner. The Giambullari case is documented elsewhere \cite{jurczak2026lightonocr}; the Ausonius case is a 211-page transcription of the Aldine 1517 \emph{Opera}. The remaining six corpus texts are sourced from validated original-language transcriptions in open repositories (see Appendix~\ref{app:corpus}); smartphone photography plus fine-tuned VLM-OCR reduces the infrastructure barrier to a level accessible to private collectors who own period editions but lack institutional digitisation facilities.

Second, each unit is encoded as a 1024-dimensional vector using \texttt{intfloat/\allowbreak multilingual-\allowbreak e5-\allowbreak large-\allowbreak instruct} \cite{wang2024multilinguale5instruct}, the top-ranked multilingual model on the MMTEB semantic-similarity benchmark \cite{enevoldsen2025mmteb}. It handles Latin, Italian, and French without language-specific tuning. Third, pairwise cosine similarities are computed. Fourth, an edge is placed between units $i$ and $j$ if $j$ is among the $k$ nearest neighbours of $i$ (or vice versa) and their cosine similarity exceeds a threshold $t$. We call the resulting object $G_{\text{text}}$.

Three distinct graphs appear throughout this paper, and we keep them notationally separate. $G_{\text{text}}$ is the semantic $k$-NN graph just defined: one node per textual unit, edges from cosine adjacency. $G_{\text{reg}}$ is the physical \emph{register graph} realised on the atom array after embedding (atom positions placed in 2D, bilayer, or 3D by the simulated-annealing procedure of \S\ref{sec:beyond2d}; edges = pairs within the Rydberg blockade radius $R_b$). $G_{\text{design}}$ is the engineered \emph{target} graph for the QOuLiPo texts of \S\ref{sec:qoulipo}, specified by construction, of which the prose is a literary realisation. For natural texts $G_{\text{text}}$ is the canonical graph and $G_{\text{reg}}$ approximates it lossily (Table~\ref{tab:2d_2L_3d_ladder}). For engineered exact-UDG and exact unit-ball-graph (UBG; the bilayer and 3D analogues of UDG) texts (\emph{Le venticinque stanze}, \emph{Il castello}, \emph{Pascal/M\'enil}, \emph{The Friar's Notebook}, plus the Archimedean and king-grid families of Table~\ref{tab:oulipo_texts}) $G_{\text{design}} = G_{\text{reg}}$ by construction, and the MIS benchmark is $G_{\text{design}}$ itself. For engineered $k$-NN texts (\emph{Nithard's Wager}, \emph{Le Pari de Nithard~II/III}, etc.) $G_{\text{design}}$ is the pre-prose target and is the canonical graph on which the QPU is benchmarked.

But if the pipeline starts with a neural embedder, isn't every structural claim hostage to that embedder? We address this in Appendix~\ref{app:embedder_validation}, running four natural texts (Dante, Boethius, Galileo, Giambullari) through three competitive multilingual transformers (\texttt{e5-large-instruct}, \texttt{bge-m3}, \texttt{arctic-embed-l-v2}) and highlight two findings. The backbone ratio $\text{MIS}/N$ is stable across all three: backbone \emph{size} is an embedder-robust property of the text. The rigidity $\rho$ is not: the three models disagree on which corpus is most rigid, so rigidity should be read as a property of the (text, embedding) pair. The engineered texts of \S\ref{sec:qoulipo}, whose graph structure is fixed by construction, serve as the calibration anchor that makes the rigidity metric interpretable across the corpus.

The pipeline has three free parameters: $N$ (the number of units; our natural-text corpus spans $N = 27$ for Ausonius's \emph{Epigrammata} up to $N = 151$ for Giambullari's full treatise), $k$ (the $k$-NN parameter controlling graph density; default 8 or 16), and a cosine similarity threshold $t = 0.78$ (almost never binding because $k$ dominates).

In stage 5 of Figure~\ref{fig:giambullari_workflow} the 2D histogram is the QPU output and the 3D histogram is produced by a classical \emph{emulator}, since no current processor accepts 3D atom layouts. More generally, the emulator is the workhorse of every step of the pipeline that precedes a QPU shot. The emulator is a GPU-based numerical simulation that integrates the same Rydberg Hamiltonian the physical processor would implement, producing the bitstring distribution the real device would output --- either as ideal noiseless dynamics or with calibrated hardware-noise models (laser dephasing, SPAM, finite-temperature broadening) when closer-to-device predictions are needed. Our runs use Pasqal Cloud's tensor-network backend \texttt{EMU\_MPS} \cite{pasqal2024emumps}, which represents the quantum state as a matrix product state \cite{schollwock2011dmrg} and is efficient as long as the transient entanglement stays below a bond-dimension ceiling the GPU can hold in memory. Most runs were performed on Pasqal Cloud's A100 back-end, with a subset of larger instances run on a separate cloud platform with H100/H200 GPUs. The practical scaling limit is set by bond-dimension growth and depends on $N$, pulse duration $T$, bond dimension $\chi$, and graph density; for the adiabatic pulses used in this paper, Vovrosh \emph{et al.} \cite{vovrosh2025crossover} locate the exponential-cost crossover in the $N \approx 100$--$150$ range, with engineered low-entanglement UDG instances reaching further before the back-end stops converging. This is the regime where direct classical simulation of the quantum dynamics, rather than classical solution of MIS itself, becomes resource-limited. In practice we use the emulator to develop and debug the workflow on each instance, then run a subset of the debugged and optimised cases on the QPU.

Throughout the rest of this paper we use \emph{QPU} as shorthand for FRESNEL\_CAN1 and \emph{EMU} for EMU\_MPS. Marker shapes and colours in the figures follow the same convention.

% ─── Section: Backbone Fidelity ──────────────────────────────────────────

\section{The Structural Backbone}\label{sec:backbone}

\begin{figure}[!b]
\centering
\includegraphics[width=1\textwidth]{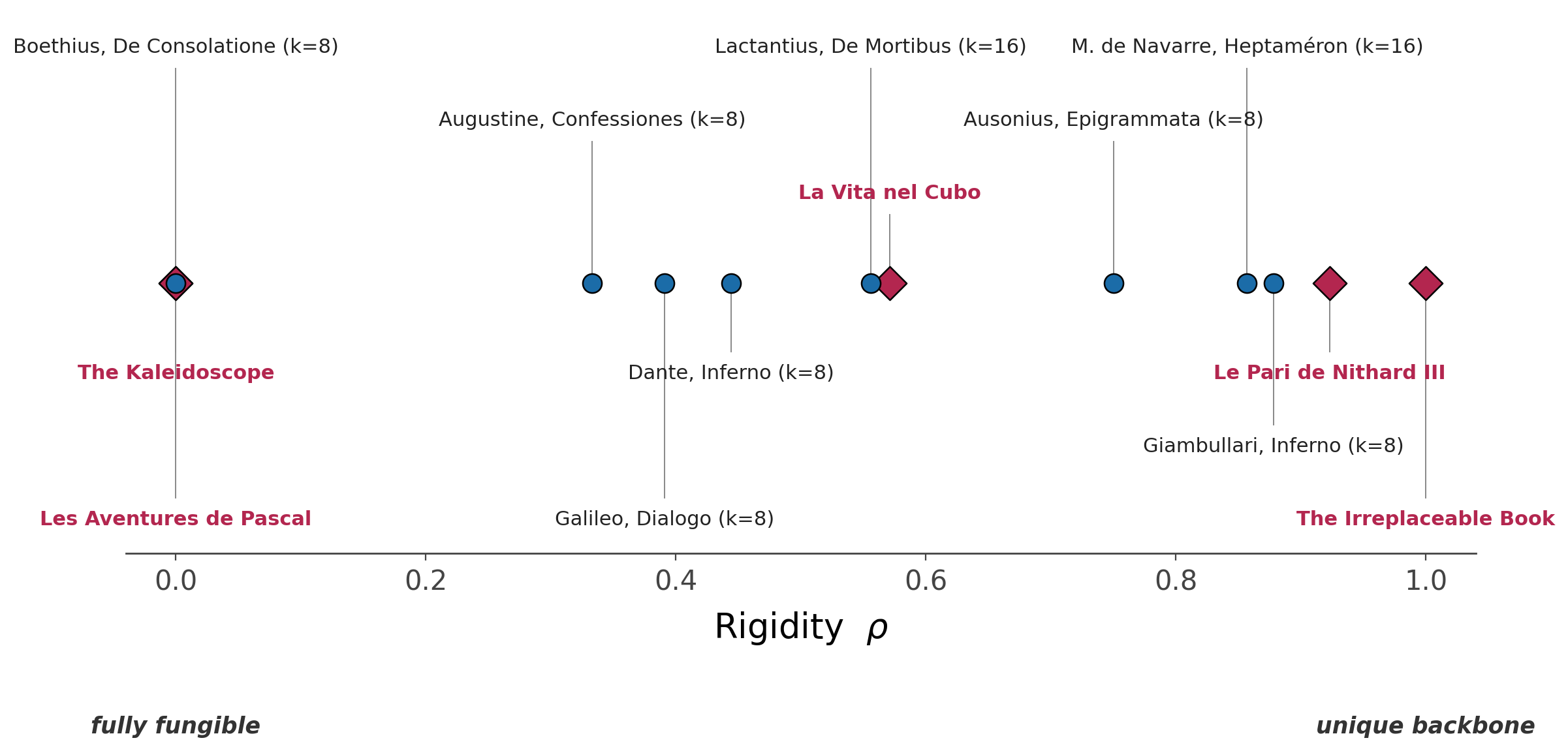}
\caption{The corpus laid out along the rigidity axis $\rho \in [0, 1]$, between the \emph{fully fungible} limit ($\rho = 0$, every selected unit has a structural substitute) and the \emph{unique backbone} limit ($\rho = 1$, every selected unit is irreplaceable). \emph{Blue circles}: natural texts (at the $k_\text{nn}$ each text is most discriminating at; see Table~\ref{tab:dh_summary}). \emph{Red diamonds}: engineered QOuLiPo texts written for this paper.}
\label{fig:rigidity_spectrum}
\end{figure}

Two natural questions arise about the MIS backbone of a text: how big it is, and how unique it is. The first is uninformative --- backbone size $\mathrm{MIS}/N$ stays in a narrow $0.10$--$0.30$ band across the corpus. The second is what discriminates between texts, and we call it \emph{rigidity}.

Rigidity $\rho$ is the fraction of MIS nodes that appear in every optimal solution. $\rho = 1.0$: unique backbone, every selected unit irreplaceable. $\rho = 0.0$: maximally degenerate, every selected unit substitutable. The underlying object --- the intersection of all optimal solutions, which we call the \emph{persistent core} (the standard \emph{persistency set} of integer programming \cite{nemhauser1975vertexpack,hammer1984persistency,boros2002pbo}) --- is standard in combinatorial optimisation, and in the Ising/Rydberg picture of \S\ref{sec:intro} it is precisely the degeneracy of the MIS ground state: $1 - \rho$ counts how many backbone slots the ground-state manifold can swap between optima. What is new is applying it to literary texts as a structural metric. Figure~\ref{fig:rigidity_spectrum} plots the spectrum across the corpus.

We believe the rigidity spectrum exposes structural properties that word-frequency methods such as Voyant Tools \cite{rockwell2016hermeneutica} and topic models do not directly access. Dante's \emph{Inferno} sits at a moderate $\rho$ because the terzina structure creates local substitutability while the cosmological progression anchors the overall arc; Galileo's \emph{Dialogo} sits low because each day's discussion is relatively self-contained, so modular reshuffling preserves the argumentative structure; the \emph{Heptam\'eron} at $k = 16$ rises because of a hard core of nouvelles distributed across the frame days that no optimal backbone can omit. Boethius's \emph{De Consolatione} at the opposite extreme raises a substantive question (whether the structural fungibility reflects the philosophical argument's internal coherence or simply the text's topical homogeneity at the embedding level) which we leave to Boethius specialists. The numerical values and the table-format readings follow in Table~\ref{tab:dh_summary}.

\subsection{What does a MIS backbone actually look like? An example}\label{subsec:worked_example}

It is worth showing the pipeline once on a text small enough to print in full. We wrote a twelve-paragraph English mini-text, \emph{Twelve Hours of a Coastal Town}, with a four-cluster structure: paragraphs 1--3 are the dawn departure, 4--6 the mid-morning market, 7--9 the silent noon, 10--12 the late-afternoon return. Each paragraph is embedded by \texttt{e5-large-instruct} and connected to its $k = 3$ nearest neighbours. The MIS is computed by ILP. The resulting graph has $N = 12$ vertices, $E = 26$ edges, and density $d = 2E/(N(N-1)) = 0.39$. The MIS has size 4, and one of the nine optimal solutions is $\{3, 4, 7, 10\}$, exactly one from each cluster. The full text is in Appendix~\ref{app:twelve_hours}; here are the four selected paragraphs:
\begin{quote}\small
\textbf{(¶3)} \emph{From the cliff above the beach an old fisherman watches the two crews vanish. He no longer goes to sea. He keeps a small notebook of weather signs and writes one line each morning, always about wind direction and the colour of the horizon.}\\[2pt]
\textbf{(¶4)} \emph{By mid-morning the market square is loud with bargaining. Fishwives shout the prices of mackerel and sardines, butchers slap the boards with their cleavers, the children of the town run between the stalls stealing peaches when no one is looking.}\\[2pt]
\textbf{(¶7)} \emph{After lunch the town falls silent. The shutters close one by one, the dogs sleep in the shadow of the church wall, and the only sound is the rope of a small boat creaking against its bollard down at the harbour. Even the gulls have stopped screaming.}\\[2pt]
\textbf{(¶10)} \emph{By late afternoon the tide has come in and is licking the keels of the boats hauled up that morning. The same boy who carried the bait bucket at dawn now sits on the gunwale of his father's boat, mending a torn net with a wooden needle, his fingers already as quick as a man's.}
\end{quote}
Those four paragraphs are a coherent micro-summary: a single scene from each phase of the day, no scene repeating, every theme represented. The backbone in the sense used here is the maximum independent set, which in this deliberately clustered example selects one representative from each thematic phase. The omissions are instructive: the morning departure (¶1, ¶2) collapses into ¶3, the cheese-seller and priests (¶5, ¶6) into ¶4, the cat and beekeeper (¶8, ¶9) into ¶7, the second-crew return (¶11, ¶12) into ¶10. The independence constraint forces one representative per cluster rather than the most vivid or quotable.

Compare this to popularity-weighted summarisation (TextRank, LexRank, centroid-based scoring). Those methods rank units by centrality and return the top-scoring ones. The result: they systematically pick units that \emph{cluster together}. The two most central market paragraphs both get selected while the quietest noon paragraph never makes the cut. MIS does the opposite: the independence constraint \emph{forbids} picking two similar paragraphs, forcing the backbone to spread across the thematic space. This is what we want from a structural backbone of a literary text.

The persistent core consists of the two paragraphs in every optimal solution: $\{$¶3, ¶10$\}$, the cliff-watcher at dawn and the boy mending nets at evening --- giving $\rho = 0.5$. Even when the algorithm can swap the noon and market choices, those two anchors are irreplaceable. $\rho = 0.5$ means two of the four backbone paragraphs are structural invariants; the other two have equally good substitutes.

The same operation, applied to Dante or Giambullari, performs this task at scales where the reader can no longer verify by inspection.

\begin{table}[!b]
\centering
\small
\begin{tabular}{lcccp{6.3cm}}
\toprule
\textbf{Corpus} & $k_\text{nn}$ & $\rho$ & \textbf{\# opt.} & \textbf{Literary reading} \\
\midrule
M. de Navarre                       & 16 & 0.857 & 6           & Frame narrative: a 12-nouvelle hard core spread across several frame days \\
Lactantius                         & 16 & 0.556 & 7           & Chronological narrative of the deaths of the persecuting emperors; semi-rigid spine \\
Dante              &  8 & 0.444 & 17          & The macro-descent through the circles anchors the arc; canto-to-canto transitions within a circle remain locally substitutable \\
Giambullari                        &  8 & 0.878 & 24          & Highly rigid topographical reconstruction; 36 of 41 backbone pages in every optimum \\
Galileo            &  8 & 0.391 & $\geq$\,500 & Four dialogue-days, each topically self-contained; high modularity within each day's discussion \\
Ausonius                            &  8 & 0.750 & 4           & 27 short occasional poems (satirical, epitaphic, descriptive); 6 of the 8 backbone poems irreplaceable, the MIS forcing strongly distinct backbone selections in a dense graph ($d = 0.464$) \\
Augustine               &  8 & 0.333 & 5           & Meditation on the opening of Genesis; semi-rigid, with three chapters present in every optimal backbone \\
Boethius   &  8 & 0.000 & $\geq$\,500 & Fully fungible: no single section is structurally privileged by the MIS criterion \\
\bottomrule
\end{tabular}
\caption{Per-text rigidity values across the natural-text corpus, with a one-line literary reading for each. $k_\text{nn}$ = number of nearest neighbours each node is connected to in the cosine-similarity graph; $\rho$ = rigidity (fraction of MIS nodes present in every optimal solution); \# opt.\ = number of distinct optimal MIS solutions (lower-bounded at the 500-solution enumeration cap).}
\label{tab:dh_summary}
\end{table}

\subsection{Rigidity readings across the corpus}

Rigidity is computed classically with off-the-shelf ILP, fast at our corpus sizes ($N \leq 151$): finding one optimal MIS takes well under a second on every natural text, and enumerating the optimal-MIS manifold up to a $500$-solution cap takes between a fraction of a second (\emph{Heptam\'eron} at $k{=}16$, $6$ distinct optima) and ${\sim}\,$two minutes (Boethius, where enumeration hits the cap); per-instance ILP timings are listed in Appendix~\ref{app:bench_results}, Table~\ref{tab:classical_benchmark}. For the rows where enumeration hits the 500-solution cap, we lift $\rho$ from a sampled estimate to an exact certified value via per-vertex ILP-exclusion (testing, for each backbone page $v$, whether MIS shrinks when $v$ is forced out), which removes any dependence on the cap.

Table~\ref{tab:dh_summary} collects rigidity across the corpora with a one-sentence literary reading. ``\# opt.'' is the number of distinct optimal MIS solutions (exact below the 500-solution cap, lower-bounded otherwise). Each row uses the $k_\text{nn}$ at which rigidity is maximally discriminating for that text. The literary readings in Table~\ref{tab:dh_summary} and in the rest of this section are sanity checks that the rigidity metric produces human-legible distinctions, not closed interpretations: $\rho$ is a property of the (text, embedding) pair --- a situated metric, in the sense of recent work on computational hermeneutics \cite{kommers2025hermeneutics} --- and a different embedding may reorder the rigidity ranking, as Appendix~\ref{app:embedder_validation} documents directly.

We fix \texttt{e5-large-instruct} throughout for reproducibility; under that fixed embedder, $\text{MIS}/N$ is stable across alternatives but $\rho$ is not (Appendix~\ref{app:embedder_validation}).

\subsection{Is the backbone real? Three checks}\label{sec:validation}

With the pipeline fixed, three further worries remain before reading the values of Table~\ref{tab:dh_summary} as properties of the texts themselves rather than of the apparatus: that the same numbers would come out of a graph with the right edge count alone; that they depend on which particular optimal backbone the solver happens to return; and that simpler off-the-shelf summarisation would do the same job. We address each in turn.

\emph{First, is it the semantics or just the edge count?} If we keep the number of units and the number of edges fixed, but reshuffle which unit connects to which, the rigidity disappears. Giambullari's canonical graph ($N = 151$, $E = 866$, $\mathrm{MIS} = 41$) sits far below random ensembles built on the same $(N, E)$: both Erd\H{o}s--R\'enyi random graphs (each pair of units connected with the same fixed probability; 100 trials) and configuration-model graphs (edges reshuffled while keeping each unit's number of neighbours fixed; 65 trials) yield larger MIS than the real graph, at the strongest empirical bound the trial counts allow ($p \leq 1/(n+1) \approx 0.01$ for ER, $\approx 0.015$ for the configuration model; no randomised graph in either ensemble matched the real text's rigidity). Rigidity discriminates even more sharply: the real Giambullari graph reaches certified $\rho = 0.878$, while randomised graphs of the same size do not. The identity of the edges --- the semantics of the text --- is what compresses the backbone, not the edge count. The reading would fail this check if rigidity rankings on real texts were statistically indistinguishable from those on degree-matched random graphs; none of our corpus shows that pattern.

\emph{Second, do different optimal backbones tell the same story?} The MIS algorithm typically returns one of several equally-good backbones. If the literary reading depended on which one the solver happened to find, the metric would be unstable. We sampled three of the 24 optimal backbones of Giambullari (each 41 pages) and generated a plain-text summary of each with Claude Sonnet 4.5, plus a fourth summary of the full 151-page text. Re-embedding the four summaries gives pairwise cosine similarity above $0.95$ throughout: distinct optimal backbones generate the same narrative. This is a qualitative robustness check rather than an independent validation --- the same model family writes the QOuLiPo prose and these summaries, so we read the null-model and per-vertex ILP certifications above as carrying the structural load. The check would fail if summaries from distinct optimal backbones diverged substantially.

\emph{Third, would simpler off-the-shelf methods give the same answer?} We tried two standard summarisation methods that, like MIS, pick a small subset of units ($k$-center clustering \cite{gonzalez1985kcenter,hochbaum1985kcenter} and greedy facility-location \cite{nemhauser1978submodular,krause2014submodular}) at the same subset size as our MIS, on thirteen rows of the natural-text corpus. Both overlap with our backbone only modestly (median around $30\%$ and $20\%$ respectively) and both routinely pick adjacent, semantically near-duplicate units. The reason is structural: neither method forbids picking two similar units, so neither produces a non-redundant backbone of mutually-dissimilar units. MIS does, and that is what makes the difference.

% ─── Section: OuLiPo ─────────────────────────────────────────────────────

\section{QOuLiPo: Engineering the Text Graph}\label{sec:qoulipo}

\begin{figure}[!b]
\centering
\includegraphics[width=0.95\textwidth]{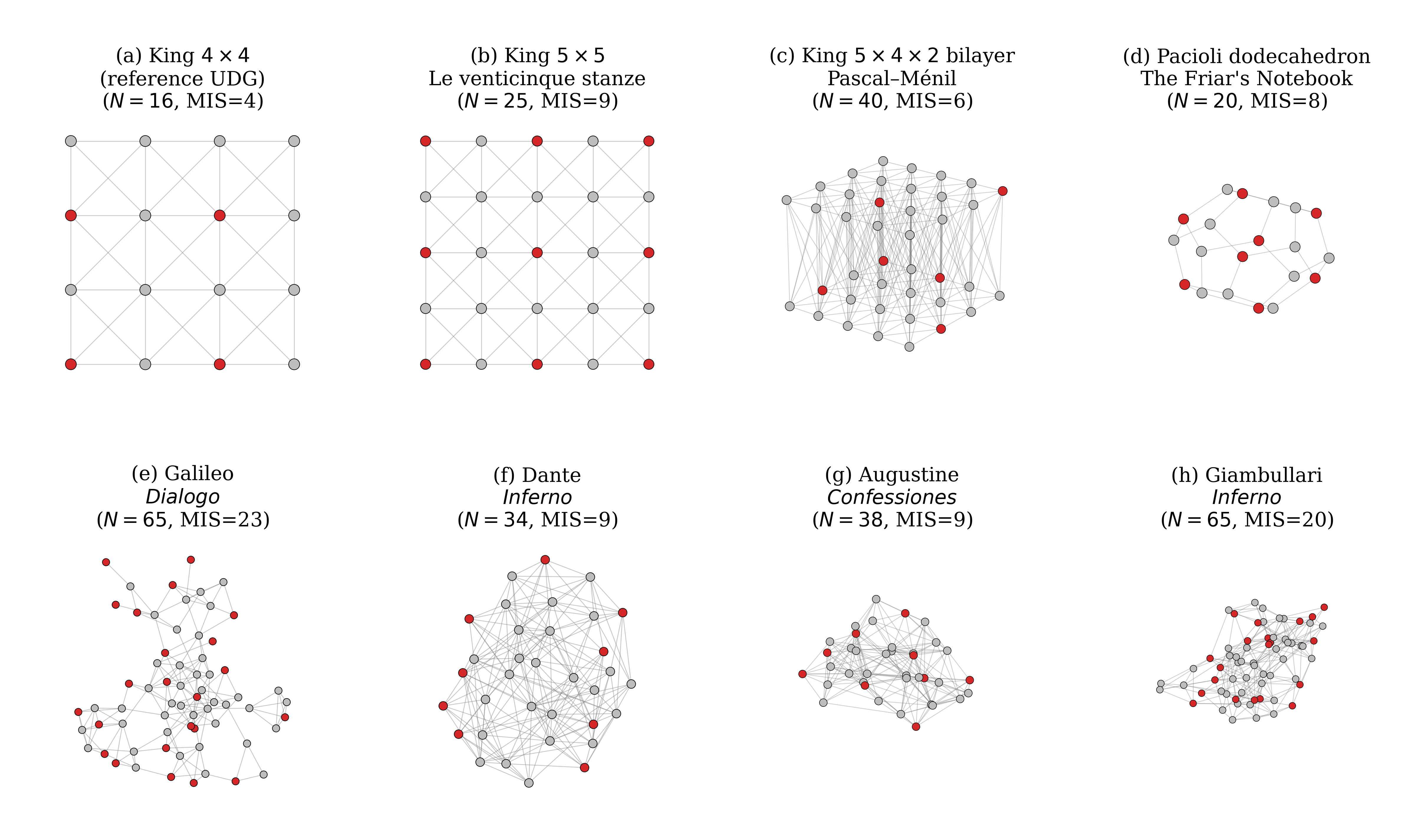}
\caption{Register targets, with the classical MIS highlighted on each graph. \emph{Top row --- QOuLiPo (engineered) texts}: graphs designed by construction as exact unit-disk or unit-ball graphs. \emph{Bottom row --- natural texts}: $k$-nearest-neighbour graphs in multilingual sentence-embedding space. Red nodes are members of an exact MIS computed by ILP.}
\label{fig:graph_gallery}
\end{figure}

\emph{QOuLiPo} (pronounced ``koolipo'') is a portmanteau of \emph{quantum} and \emph{OuLiPo}, and is introduced here for the first time. It is a constrained-writing project that treats a book's semantic graph as an object to be \emph{designed} rather than discovered, and designs it under constraints tailored to quantum hardware (with a focus on neutral-atom architectures in this paper). The result is a family of readable literary texts whose graph-theoretic invariants are controlled by construction. Figure~\ref{fig:graph_gallery} previews the visual flavour of these register targets, alongside the natural-text graphs against which they are contrasted.

\subsection{Ancestors: from Ausonius to Perec}

OuLiPo (the \emph{Ouvroir de litt\'erature potentielle}, founded in 1960 by Raymond Queneau and Fran\c{c}ois Le Lionnais \cite{oulipo1981atlas}) produces literary works under explicit formal constraints. Queneau's \emph{Cent mille milliards de po\`emes} \cite{queneau1961sonnets} combines fourteen sonnets by permutation; Perec's \emph{La Disparition} \cite{perec1969disparition} excludes the letter \emph{e}; Roubaud's \emph{$\in$} \cite{roubaud1967epsilon} treats poems as set-theoretic elements.

Where classical OuLiPo constrains letters (lipograms), verses (cento), or chapter order (knight's tours), QOuLiPo extends the tradition to the design of the graph itself: the constraint is the canonical graph $G_{\text{design}}$, and the prose is its literary realisation. The notation ($G_{\text{text}}, G_{\text{reg}}, G_{\text{design}}$) and the broad split between exact-UDG and $k$-NN-engineered texts were introduced in \S\ref{sec:pipeline} alongside the natural-text pipeline; a finer decomposition appears later in this section, and the full inventory in Table~\ref{tab:oulipo_texts}. What QOuLiPo adds beyond classical OuLiPo is that the constraint is now an object the QPU can execute on: each text is simultaneously a readable literary work and a quantum-benchmark instance whose graph structure is known by construction.

The synthesis of constrained literature with combinatorial mathematics is older than OuLiPo. In the preface to his \emph{Cento Nuptialis} (c.~370 AD), Ausonius of Bordeaux (late-Latin poet, tutor to Emperor Gratian) compares his cento technique (a wedding poem composed of recombined Virgilian half-lines) to the Stomachion, the geometric puzzle attributed to Archimedes whose 14 pieces admit 17,152 configurations \cite{netz2011palimpsest}. Ausonius writes: \emph{``like that game which the Greeks call ostomachion \ldots\ from various joinings, a thousand species are simulated''} \cite{ausonius_cento_preface}. The text we OCR'd for this work (the Aldine 1517 \emph{editio princeps}) thus contains, in its own preface, the constraint-literature $\leftrightarrow$ combinatorial-mathematics parallel this paper operationalises. Following \cite{blossier2009oulipiens}, we treat Ausonius as an ancient Oulipian. The Stomachion preface is the strongest ancient-source evidence for the claim.

A particularly striking twentieth-century precedent for us is Perec's later and more ambitious novel \emph{La Vie mode d'emploi} \cite{perec1978vie}, whose 99 chapters are ordered by an open knight's tour on a $10 \times 10$ grid representing the elevation of a Parisian apartment building. The chapter order is a Hamiltonian path on the $10 \times 10$ knight's graph; each of the 99 chapters is further overdetermined by 42 thematic elements distributed via 21 Graeco-Latin bi-squares.

What Perec's scholarship has to our knowledge not observed is that this constraint graph appears not to admit a 2D unit-disk realisation (we have not found one in the literature nor constructed one ourselves). Perec chose his constraint for literary reasons --- because it forced the narrator to skip around the building unexpectedly, and because the deliberately omitted 66th chapter is the structural counterpart of the novel's protagonist Bartlebooth's failure to complete his life-project --- but the constraint he chose is a non-planar, bounded-degree graph of exactly the kind discussed in \S\ref{sec:beyond2d}. Half a century before the neutral-atom hardware community began wrestling with non-unit-disk encodings, the most famous OuLiPo novel had already built itself on exactly such a graph.

Italo Calvino, the first Italian OuLiPo member and Perec's nearest peer in the combinatorial tradition, provides a parallel lineage of graph-structured fiction: \emph{Il castello dei destini incrociati} \cite{calvino1973castello} constructs narratives from tarot-card adjacencies; \emph{Se una notte d'inverno un viaggiatore} \cite{calvino1979notte} nests ten interrupted novels inside a frame that is itself a combinatorial traversal of narrative beginnings; and the \emph{Lezioni americane} \cite{calvino1988lezioni} (in their fifth memo on \emph{Multiplicity}) theorise the OuLiPo programme explicitly as a literature of combinatorial networks. Outside OuLiPo proper, Jorge Luis Borges (\emph{La biblioteca de Babel}, \emph{El jardín de senderos que se bifurcan}, both 1941 \cite{borges1941jardin}) extends the same combinatorial imagination into Spanish-language fiction. Our engineered texts continue this combinatorial lineage; and where Perec's and Calvino's graph structures were chosen for literary reasons (with no hardware available against which to test them), we design ours explicitly with the properties of neutral-atom processors in mind.

\subsection{The QOuLiPo design space}\label{subsec:design-space}

\begin{figure}[!t]
\centering
\includegraphics[width=0.6\textwidth]{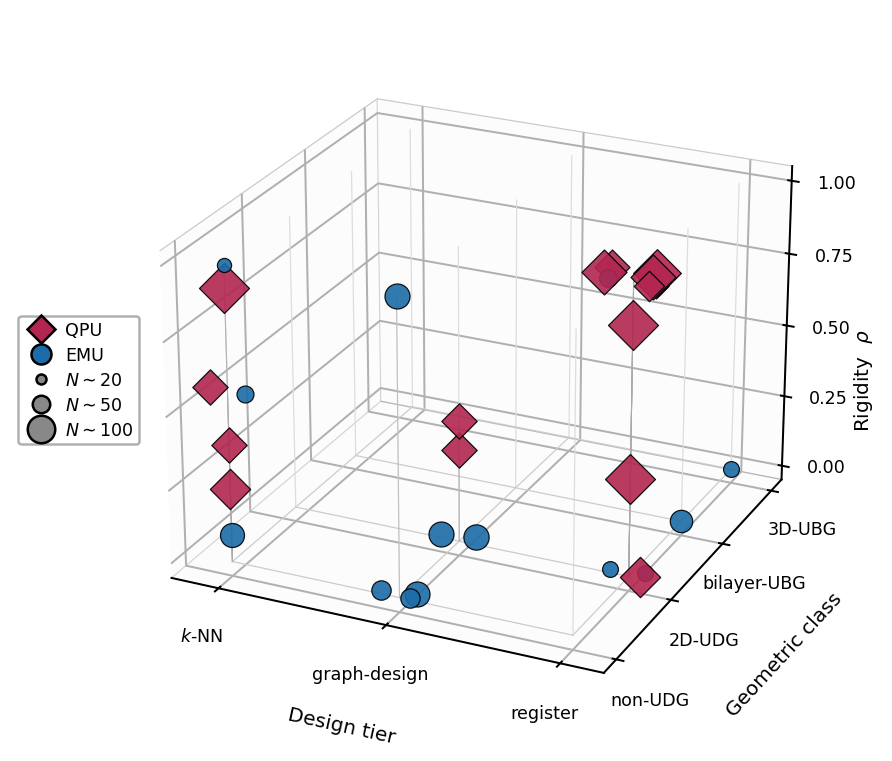}
\caption{The full QOuLiPo corpus (twenty-nine engineered texts) in the three-dimensional design space (design tier $\times$ geometric class $\times$ rigidity). Each marker is one engineered text; markers are slightly offset within each (tier, geometric-class, $\rho$) bucket so all 29 are visible. Red diamonds: corpus members run on the QPU (FRESNEL\_CAN1). Blue circles: corpus members run only on the emulator. Marker size scales with $N$.}
\label{fig:design_space}
\end{figure}

We generated twenty-nine constrained texts (full inventory in Appendix~\ref{app:oulipo_full_table}, Table~\ref{tab:oulipo_texts}). The corpus is organised along three orthogonal axes (design tier, geometric class, rigidity) and Figure~\ref{fig:design_space} shows where every text lands in the resulting three-dimensional space.

The generation mechanism produces the prescribed canonical graph $G_{\text{design}}$ and steers the embedder's $G_{\text{text}}$ toward it. Each page is assigned $m$ threads from a pool of $T$ thematic threads; pages sharing at least $\tau$ threads are the edges of $G_{\text{design}}$. The prose is then written with thread-specific keywords so that the embedder reads each prescribed pair as high cosine similarity, and $G_{\text{text}}$ approximates $G_{\text{design}}$. The $k$-NN parameter then sets the realised density of $G_{\text{text}}$ through the rule $d \approx k/N$; for high-density targeting ($d \geq 0.30$) we set $k \geq 0.30 N$.

The prose is produced as constrained generation in the OuLiPo sense: the human supplies the constraint (target graph, thread / register / poetic-form assignments, narrative voice, rigidity target, language) and Anthropic's Claude (Opus 4.6 / 4.7) generates page-level prose that satisfies it. Validation is mechanical (ILP MIS and $k$-NN edge recovery, in place of the syllable counts or vowel-avoidance of classical OuLiPo); the author validates each page against the deposited graph and iterates where the constraints are not met, with several texts going through five or more revision cycles. Two distinct quantities are worth separating here. For register-led exact-UDG and exact-UBG texts, register fidelity is $1.000$ by construction --- the geometric identity $G_{\text{design}} = G_{\text{reg}}$ holds before any prose is written. Separately, the \emph{semantic-realisation fidelity} --- the edge recall between the embedder-derived $G_{\text{text}}$ and the prescribed $G_{\text{design}}$ once the prose exists --- is a property of the writing, not the geometry. The canonical files (\texttt{source/page\_NNN.txt} in the Zenodo deposit accompanying this paper \cite{jurczak2026qoulipo_dataset}; full description in \S\ref{sec:data}) are the final versions. The corpus varies in literary ambition: some texts are pursued as serious literary objects, others kept as benchmark instances. As a first attempt, the literary pieces invite contributions from writers, poets, and OuLiPians to turn these graph-bound exercises into genuinely artistically relevant works.

The engineered texts arrange themselves along three increasingly strict design tiers. In the \emph{$k$-NN-led} regime, a target graph guides the prose generation, but the canonical graph is the embedder's $k$-NN of the final prose (\emph{Nithard's Wager}, \emph{Le Pari de Nithard~II/III}, \emph{La Vita nel Cubo}, \emph{Sonetti dal Tesseratto}, \emph{La Partition du Texte}); the constraint is on $(N, d)$ and on the rough thread structure, not on the precise edge list. In the \emph{graph-design-led} regime the canonical graph is fixed exactly --- as a $K_{25,25}$ bipartite (\emph{The Trial of Nithard}), a Sierpinski hierarchy (\emph{The Fractal Book}), 17 disjoint $K_3$ (\emph{The Kaleidoscope}), a hub-and-spoke trap for greedy heuristics (\emph{The Reader's Trap}), a planar $5{\times}10$ grid (\emph{The Map of the Text}), or a unique-MIS double-domination construction (\emph{The Irreplaceable Book}) --- but the text has no register coordinates: pages are graph nodes, and the prose realises the prescribed adjacencies. In the \emph{register-led} regime pages are placed at explicit register coordinates so that $G_{\text{design}} = G_{\text{reg}}$ holds exactly --- the king-grid families (\emph{Le venticinque stanze}, \emph{Il castello}, \emph{The Book of Eighty-One Squares}, \emph{Eighty-One Faces of a Pomegranate}, the Pascal/M\'enil pair), the centred-hex texts (\emph{Treize sur trente-sept}, \emph{Hours of the Day}), the Archimedean tilings (\emph{Kagome of Late Summer}, \emph{The Postman's Songbook}), \emph{The Incarnate Graph} (random UDG placement), and the bilayer and three-dimensional extensions (\emph{Pascal--M\'enil bilayer}, \emph{The Friar's Notebook}).

A second axis cuts across this tier ladder: the geometric class of the canonical graph. The $k$-NN-led family is non-UDG by construction, since the canonical graph is a 1024-dimensional cosine $k$-NN, not a unit-disk graph. The register-led family is 2D-UDG, or bilayer/3D-UBG for the bilayer and dodecahedron texts, by construction. The graph-design-led family is mixed: most members are 2D-UDG-realisable (\emph{The Map of the Text} as a planar grid, \emph{The Reader's Trap} as a hub-spoke star, \emph{The Kaleidoscope} as 17 well-separated triangles, \emph{The Fractal Book} as a Sierpinski hierarchy at moderate density), but three appear non-UDG: the $K_{25,25}$ bipartite of \emph{The Trial of Nithard} (which contains $K_{4,4}$ and is therefore not unit-disk by a standard kissing-number argument), the double-domination construction of \emph{The Irreplaceable Book}, and \emph{Les Jumeaux du Graphe} ($C_{30}$ plus eight same-parity chords) all defeated our embedding attempts; we have not produced formal non-embeddability proofs for the latter two.

The third dimension of Figure~\ref{fig:design_space} is rigidity $\rho$. The corpus is designed to cover the full $[0, 1]$ interval, with rigid backbones, maximally degenerate constructions, and intermediate values all represented.

\begin{figure}[!t]
\centering
\includegraphics[width=0.95\textwidth]{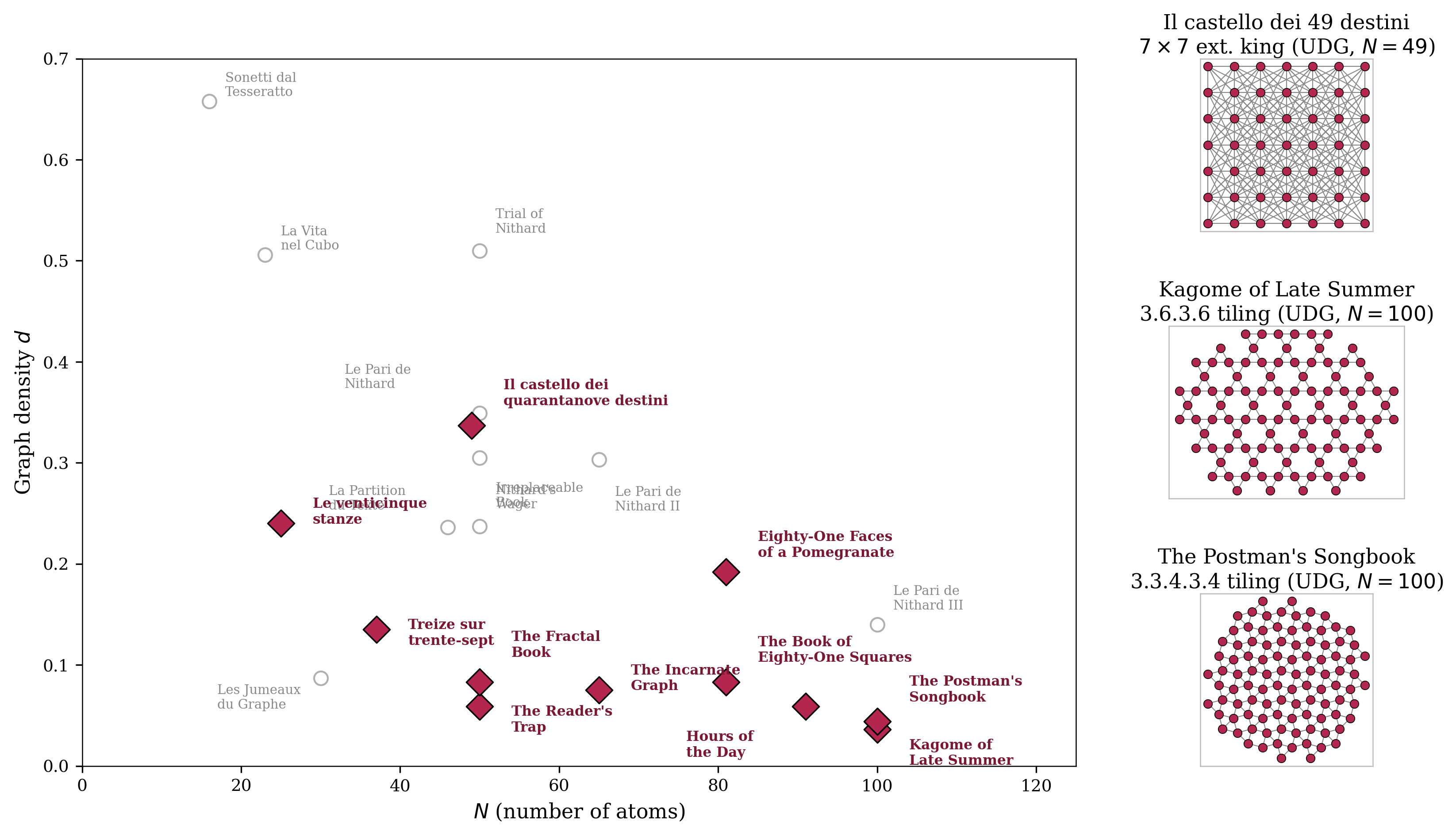}
\caption{Engineered QOuLiPo corpus on the $(N, d)$ plane. \emph{Filled red diamonds}: 2D-UDG texts run on the QPU. \emph{Open circles}: non-UDG engineered texts ($k$-NN-led and graph-design-led). Bilayer / 3D extensions and EMU-only 2D-UDG texts are not shown; see Table~\ref{tab:oulipo_texts} for the full inventory.}
\label{fig:cazals}
\end{figure}

Figure~\ref{fig:cazals} positions the engineered QOuLiPo corpus on the $(N, d)$ design plane, where each graph occupies a point defined by the register size $N$ (one atom per vertex) and the graph-theoretic density $d = 2|E|/(N(N-1))$. These two quantities decide which constraints the graph stresses on the QPU. $N$ sets the hardware budget: FRESNEL\_CAN1 supports up to 100 atoms, so the texts that exercise the QPU most are those that hit the upper end of that budget. Density $d$ controls how constrained the MIS problem is: sparse graphs have many degenerate independent-set configurations and are trivially solved by any heuristic, while dense graphs have small MIS, larger branching factors, and empirically slower classical runtimes; Cazals \emph{et al.}\ \cite{cazals2025mis} document this trend systematically on UDG instances at full triangular-lattice filling and $N \geq 50$. Engineering the QOuLiPo corpus to span the $(N, d)$ plane was therefore an explicit design goal. QPU outcomes are reported in \S\ref{sec:quantum}.

\subsection{Beyond 2D: bilayer and 3D text designs}\label{subsec:beyond2d-texts}

The QOuLiPo corpus also includes two texts whose graph structure requires \emph{more} than a planar register. \emph{The Friar's Notebook} ($N = 20$) is twenty short pieces, each in a different literary register (confession, ledger, sonnet, letter to Leonardo, recipe, dream, sermon, dialogue, prayer, epitaph, \ldots), apocryphally attributed to the Renaissance mathematician Luca Pacioli (1447--1517), whose \emph{De divina proportione} was illustrated by Leonardo, and arranged at the twenty vertices of a regular dodecahedron --- a 3D unit-ball graph that is provably not realisable as a 2D UDG or even a bilayer graph. It is the simplest QOuLiPo text that \emph{requires} full 3D.

The second is a two-layer engineered text written specifically for a bilayer register geometry. \emph{Les Aventures de Pascal} is a twenty-page apocryphal French novel about Blaise Pascal (the philosopher and mathematician), organised on a $5 \times 4$ grid whose axes are the five eras of Pascal's life (Clermont childhood, Paris salons, vacuum experiments, Port-Royal, posthumous fortunes) and four cardinal threads (reason, wager, vacuum, mystic). Each page sits at one (era, thread) cell and carries the corresponding thematic markers so that king-adjacent pages (the eight grid neighbours --- orthogonal and diagonal, after the chess king's move) share vocabulary. \emph{Scholia de M\'enil}, a parallel second text, is 20 pages of fictional 19th-century scholarly annotation by ``Adolphe M\'enil'', archivist at the Biblioth\`eque Nationale in 1878, laid out on an identical $5 \times 4$ grid whose columns are four editorial concerns (philology, biography, science-history, theology). Together the two texts form a 40-page bilayer whose graph structure --- a $5 \times 4 \times 2$ king lattice --- is fixed by construction. The self-reference is deliberate: Pascal is being read on Pasqal.

The physical realisation of both the dodecahedron and the bilayer is taken up in \S\ref{sec:beyond2d}, where they serve as engineered anchors for the natural-text 3D and bilayer experiments.

% ─── Section: Quantum Results ──────────────────────────────────────────

\section{Quantum MIS on FRESNEL QPU}\label{sec:quantum}

The text graphs of \S\ref{sec:backbone} and the engineered constructions of \S\ref{sec:qoulipo} are now loaded onto the QPU --- one atom per textual unit (page, canto, novella, paragraph, chapter --- whatever the author's segmentation gave us in \S\ref{sec:pipeline}), with one Rydberg blockade per edge of the realised register graph $G_{\text{reg}}$ --- exact for the engineered-UDG rows where $G_{\text{design}} = G_{\text{reg}}$ by construction, lossy for the natural-text rows where $G_{\text{reg}}$ approximates $G_{\text{text}}$ through the SA embedder. The MIS that ILP computes on $G_{\text{reg}}$ is the same MIS that the array's quantum ground state encodes in matter; the rest of this section reports what the hardware actually returns, on both halves of the corpus.

\begin{figure}[!b]
\centering
\includegraphics[width=\textwidth]{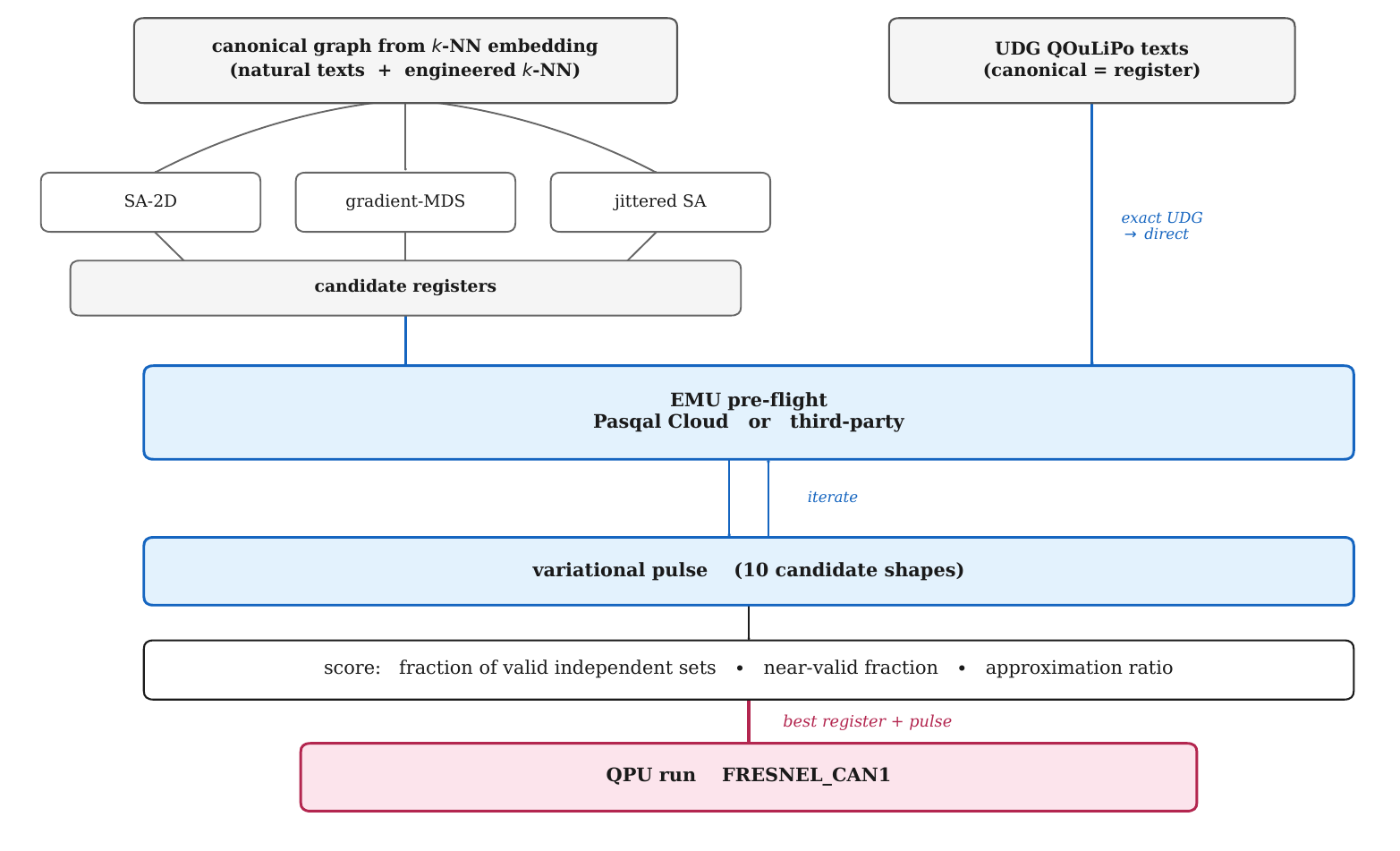}
\caption{Optimisation toolkit: the EMU $\leftrightarrow$ QPU loop. The canonical graph (text $k$-NN or designed) feeds candidate (register, pulse) pairs through EMU pre-flight; the best pair is submitted to the QPU.}
\label{fig:qpu_pipeline}
\end{figure}

\subsection{Setup: register and pulse}\label{subsec:fresnel_setup}

We implement the text-graph MIS pipeline on Pasqal's FRESNEL\_CAN1 (100-qubit, 2D rubidium register, hosted at the DistriQ quantum innovation zone in Sherbrooke, Qu\'ebec) in \emph{analog} mode: the physical interactions \emph{are} the computation, no circuit is compiled, no gates are applied --- Feynman's 1982 vision \cite{feynman1982simulating} of computing with matter, made concrete by the Rydberg blockade introduced in \S\ref{sec:intro}. All natural texts fit the QPU's register at their natural granularity (Dante \emph{Inferno} $N{=}34$, Augustine $N{=}38$, Lactantius $N{=}52$, Giambullari $N{=}65$ at the central-chapter abridgement of the $N{=}151$ full treatise, Galileo $N{=}65$; see Appendix~\ref{app:corpus}); for Dante we report QPU results on the \emph{Inferno} at $N{=}34$ (canto level) and a separate full-\emph{Divina Commedia} run at $N{=}100$ cantos, both in Table~\ref{tab:qpu_results}. QOuLiPo texts use the full 100-qubit capacity. Field of view ($\leq 46\,\mu$m radius) and minimum spacing ($\geq 5\,\mu$m) impose geometric constraints, all satisfied by the graphs of Table~\ref{tab:qpu_results}.

Figure~\ref{fig:qpu_pipeline} summarises the EMU\,$\leftrightarrow$\,QPU optimisation loop used throughout this section. EMU is a faithful proxy of the QPU in its convergent regime; where it does not converge (several large-$N$ runs in this campaign hit the bond-dimension ceiling described in \S\ref{sec:pipeline}), we transfer design rules from small-$N$ topology-matched controls to their large-$N$ targets, the \emph{Treize sur trente-sept}\,/\,\emph{Hours of the Day} pair being one such transfer.

Register engineering (the simulated-annealing fit of $G_{\text{text}}$ to FRESNEL's triangular lattice, maximising \emph{edge recall} --- the fraction of canonical $G_{\text{text}}$ edges that survive within $R_b$ on the embedded register) and pulse choice (the standard baseline ramp $\Omega = 3.3$~rad/$\mu$s, $T = 4\,\mu$s, extended to $T = 6\,\mu$s at $N \geq 81$) are documented in detail in Appendix~\ref{app:hardware_detail}. Two facts matter for what follows. \emph{For natural texts}, the SA edge-recall ceiling caps below $0.5$ at $N \geq 30$ in 2D (quantified in \S\ref{sec:beyond2d}, Table~\ref{tab:2d_2L_3d_ladder}), and the dense graphs one would most want the QPU to recover are precisely those that lose the most information at the embedding stage. \emph{For engineered exact-UDG texts}, the register is constructed so that edge recall is $1.0$ exactly --- $G_{\text{design}} = G_{\text{reg}}$ by construction --- and any deviation from ratio $1.0$ on QPU is attributable to hardware noise alone, with neither embedder mismatch nor register-fidelity loss in the way.

Where an SA-embedded or exact-UDG instance nevertheless fails on hardware (zero valid IS), the practical remedy is a tighter blockade margin $r_{\text{nnn}}/R_b$ between the smallest non-edge distance and the blockade radius, achieved by re-running SA with all non-edges pushed comfortably above $R_b$; this single lever lifts several rows of Table~\ref{tab:qpu_results} from $0/1000$ valid IS to ratios in the $0.84$--$1.00$ band, and is the largest hardware-side margin documented in this paper. Pulse engineering, by comparison, is a second-order refinement: EMU-selected variational pulse shapes improve \texttt{valid IS\%} by $+3$--$6$ percentage points over the baseline ramp on the engineered exact-UDG rows of Table~\ref{tab:qpu_results}, with the approximation ratio pinned at $1.000$ (full sweep in Appendix~\ref{app:hardware_detail}).

\subsection{QPU results}\label{subsec:qpu_results}

\begin{table}[!b]
\centering
\small
\begin{tabular}{lcccccc}
\toprule
\textbf{Text} & \textbf{$N$} & \textbf{$k$} & \textbf{$d$} & \textbf{Best-shot ratio} & \textbf{Valid IS\%} & \textbf{Near-valid\%} \\
\midrule
\multicolumn{7}{l}{\emph{Natural k-NN texts (SA-embedded)}} \\
Ausonius \emph{Epigrammata}  & 27 & 8  & 0.464 & 1.000 & 10.0\% & 36.2\% \\
Augustine \emph{Confessiones} (paragraph) & 38 & 8  & 0.293 & 1.000 & 13.5\% & 4.9\% \\
Lactantius \emph{De mortibus persecutorum} & 52 & 8  & 0.232 & 1.000 & 61.8\% & 33.0\% \\
Galileo \emph{Dialogo} & 65 & 8  & 0.079 & 0.966 & 55.9\% & 31.5\% \\
Boethius \emph{Consolatio} & 72 & 8  & 0.146 & 0.968 & 46.4\% & 2.4\% \\
M.~de Navarre \emph{Heptam\'eron} & 72 & 8  & 0.165 & 0.969 & 51.9\% & 16.4\% \\
Dante \emph{Divina Commedia} & 100 & 8  & 0.121 & 0.917 & 43.7\% & 1.7\% \\
\midrule
\multicolumn{7}{l}{\emph{Engineered k-NN texts (SA-embedded, variational pulse)}} \\
Nithard's Wager       & 50 & 16 & 0.305 & 1.000 & 59.5\% & --- \\
Le Pari de Nithard    & 50 & 16 & 0.349 & 1.000 & 60.6\% & --- \\
\midrule
\multicolumn{7}{l}{\emph{Exact-UDG and designed-graph engineered texts}} \\
Le venticinque stanze             & 25 & --- & 0.240 & 1.000 & 53.1\% & 85.3\% \\
Treize sur trente-sept & 37 & --- & 0.135 & 0.923 & 0.7\% & 8.6\% \\
Il castello dei quarantanove destini$^{\ddagger}$ & 49 & --- & 0.133 & 0.938 & 24.1\% & 62.1\% \\
The Reader's Trap & 50 & --- & 0.059 & 1.000 & 65.4\% & 93.9\% \\
The Fractal Book & 50 & --- & 0.083 & 0.947 & 60.5\% & 92.9\% \\
The Incarnate Graph & 65 & --- & 0.075 & 0.955 & 3.2\% & 27.1\% \\
The Book of Eighty-One Squares$^{\S}$ & 81 & --- & 0.083 & 0.840 & 2.9\% & 18.2\% \\
Hours of the Day$^{\S}$ & 91 & --- & 0.059 & 0.839 & 23.9\% & 62.4\% \\
Kagome of Late Summer$^{\S}$ & 100 & --- & 0.036 & 0.923 & 18.4\% & 62.3\% \\
The Postman's Songbook$^{\S}$ & 100 & --- & 0.044 & 0.861 & 30.3\% & 73.5\% \\
\bottomrule
\end{tabular}
\caption{QPU results across the corpus. \texttt{Best-shot ratio} $r$ is the largest valid-IS shot size divided by the ILP-exact MIS of the benchmark graph (the realised register graph $G_{\text{reg}}$ for natural-text and $k$-NN engineered rows; $G_{\text{design}} = G_{\text{reg}}$ by construction for exact-UDG and design-led engineered rows). \texttt{Valid IS\%} is the fraction of shots forming a valid IS. \texttt{Near-valid\%} is the regime-dependent occupancy diagnostic defined in \S\ref{subsec:qpu_results}. $^{\S}$ Pulse duration $T = 6~\mu$s; other rows $T = 4~\mu$s. $^{\ddagger}$ King-only QPU subgraph of the extended-king construction; see \S\ref{subsec:qpu_results}.}
\label{tab:qpu_results}
\end{table}

Table~\ref{tab:qpu_results} reports data for 19 QPU submissions, all at 1000 shots per row except \emph{The Incarnate Graph} at $402$ shots, organised in three blocks: natural $k$-NN texts (SA-embedded), engineered $k$-NN texts (also SA-embedded), and exact-UDG / designed-graph engineered rows. The best-shot ratio is $r = \max_{s\,\text{valid}} |s|\,/\,\alpha(G)$, where the maximum runs over the 1000 shots forming a valid IS on the benchmark graph, $|s|$ is the IS size, and $\alpha(G)$ is the ILP-exact MIS; rows with no valid shot are reported as $r = 0$. Near-valid\% on SA-embedded rows is the fraction of shots with $|w - \mathrm{MIS}| \leq 1$ (near-MIS occupation indicator); on exact-UDG rows it is the fraction of shots violating at most two canonical edges (soft-blockade tail indicator); the two are not interchangeable.

On natural-text rows the pipeline cascades through two upstream stages before any QPU shot: a $1024$-dimensional cosine $k$-NN graph built on the embedded text units, then a simulated-annealing fit of that adjacency onto FRESNEL's triangular lattice. Table~\ref{tab:qpu_results} reports for each row the approximation ratio against the realised register MIS (the standard convention in the neutral-atom MIS literature \cite{ebadi2022quantum,cazals2025mis}). On every bigger natural-text row we ran --- Augustine, Lactantius, Galileo, Marguerite de Navarre, Boethius, and Dante ($N = 38$ to $100$) --- this register-MIS ratio is $\geq 0.92$ at $14$--$60\%$ strict valid-IS rates; on the engineered $k$-NN rows with EMU-selected variational pulses it is $1.000$ at $\geq 60\%$ valid IS. The QPU delivers on the graph it is sent.

What the literary user receives is one step removed. The relevant quantity is the canonical text-MIS ratio --- the fraction of the actual backbone the pipeline returns end-to-end --- bounded above by the SA recall of the realised register. In practice, with the simulated-annealing embedder used throughout this paper, we observe that recall caps below $0.5$ at $N \geq 30$; Galileo's \emph{Dialogo} reaches that ceiling exactly and accordingly delivers the highest canonical recovery on hardware: $14$ of the $23$ backbone pages, ratio $0.61$. The near-valid Hamming-weight column in Table~\ref{tab:qpu_results} carries a complementary signal: on Galileo, $31.5\%$ of shots have Hamming weight within one of the canonical MIS size, and on Ausonius $36.2\%$ --- the QPU output clusters tightly around the right occupation regime even where individual shots fail strict independence, giving a post-processable structural signal where the strict-valid rate alone would say nothing. The upstream embedding step is the bottleneck on canonical-text MIS recovery, not the processor; closing that gap with better embedders --- gradient-MDS, SDP relaxations, learned register placement, or the 3D escape of \S\ref{sec:beyond2d} --- is a natural follow-up.

\begin{figure}[!htbp]
\centering
\includegraphics[width=0.85\textwidth]{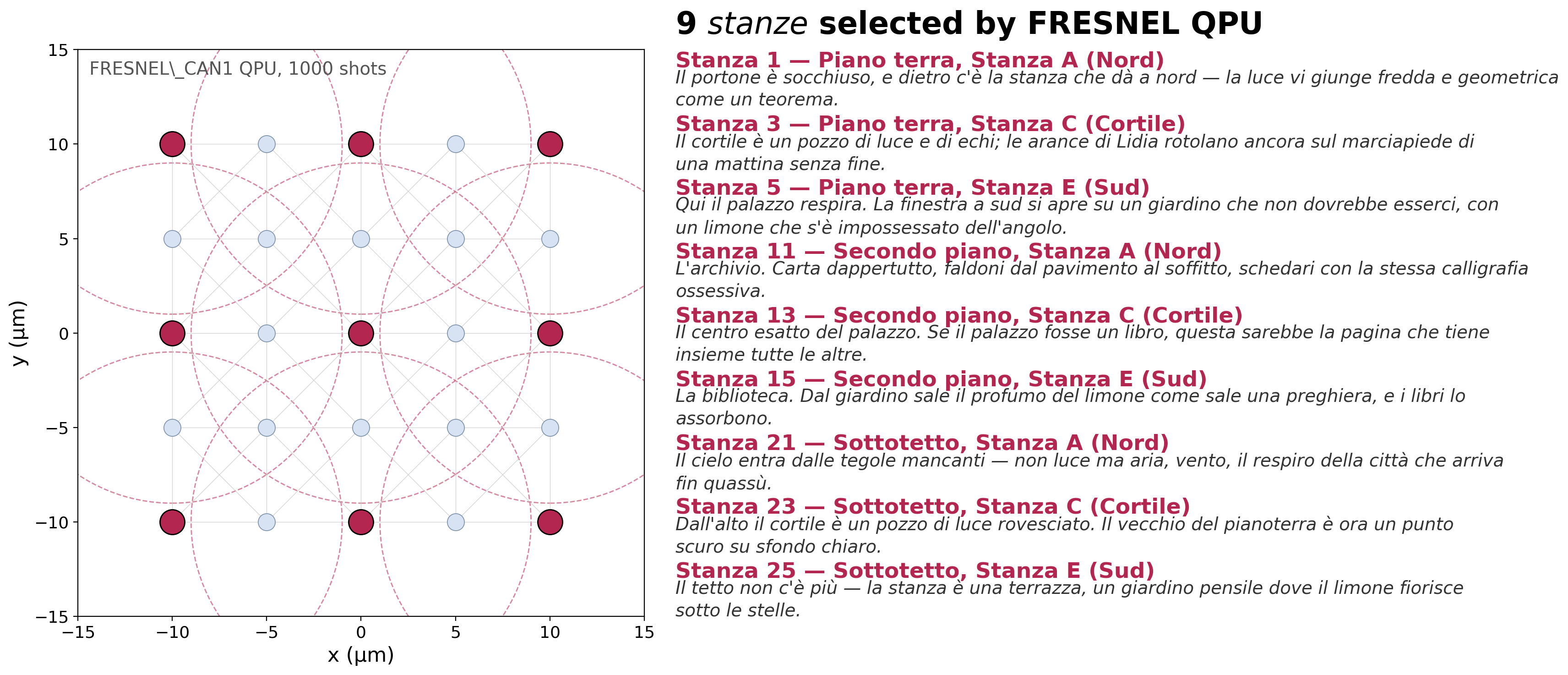}
\caption{The quantum backbone of \emph{Le venticinque stanze}. Left: register; the 9 atoms of the best valid independent set are in red. Right: the 9 selected \emph{stanze}, spanning all three floors of the building (\emph{piano terra}, \emph{secondo piano}, \emph{sottotetto}) and three positions across the building's depth (\emph{nord}, \emph{cortile}, \emph{sud}).}
\label{fig:qpu_text_case}
\end{figure}

On the exact-UDG and designed-graph engineered rows the canonical graph is by construction realised by the register, and the QPU does what the physics promises: on the cleanest instance, \emph{Le venticinque stanze} ($N=25$, an exact $5{\times}5$ king-UDG graph; the same row shown in Figure~\ref{fig:qpu_text_case}), the best QPU shot reaches the classical $9$-page MIS at ratio $1.000$, with $53.1\%$ valid independent sets across $1000$ shots. The exact-MIS shot fraction is $2.4\%$; the remaining valid shots cover sub-MIS independent sets across the structural backbone. On these rows the Rydberg blockade \emph{is} the independence constraint, with no compilation in between. The hardware exposes per-atom occupation, near-valid Hamming-weight statistics, and the full shot histogram natively in microseconds, complementing rather than duplicating the single optimum a classical ILP solver returns.

Two exact-UDG constructions in our campaign sit at FRESNEL's parameter edge and are reported here as hardware-boundary diagnostics rather than as part of the headline $0.84$--$1.00$ band of Table~\ref{tab:qpu_results}. \emph{Il castello} (extended-king on a $7\!\times\!7$ grid) would require $R_b > 14.14\,\mu$m at the standard spacing, which corresponds to $\Omega < 0.11\,\text{rad}/\mu$s --- below FRESNEL's amplitude floor $\Omega \geq \pi/10 \approx 0.314\,\text{rad}/\mu$s; the row in Table~\ref{tab:qpu_results} is the king-only subgraph (Chebyshev $\leq 1$, $R_b = 8\,\mu$m, $\mathrm{MIS} = 16$) at ratio $0.938$. \emph{Eighty-One Faces of a Pomegranate} ($N{=}81$, multi-distance $\sqrt{5}$ king) goes the other way: at $R_b = 11.5\,\mu$m the $\sqrt{5}$-distance pairs sit at $U/\Omega \approx 1.4$ --- the energy gap between the blockaded $|rr\rangle$ state and the singly-excited level is only marginally above the Rabi drive, a physics constraint that no pulse shape fixes --- and the QPU returns only ratio $0.615$. The natural remedy here is not dimensional escape (\S\ref{sec:beyond2d}) but \emph{per-atom local detuning} --- on Pasqal's hardware roadmap --- or ancilla-augmented embedding \cite{dalyac2022embed3d}. A wider Rabi-frequency window on the device, together with local-control extensions, would together extend the class of canonical graphs realisable as exact UDG; on this axis, hardware flexibility buys more useful instances than additional qubits.

Where the QPU comes closest to literally reading the book is on Galileo's \emph{Dialogo}: the SA register loses about half the canonical edges, yet the QPU still recovers $14$ of the $23$ pages of Galileo's structural backbone --- a non-trivial subset of the four-day dialogue, anchored by the most distinctive astronomical demonstrations on each day. Figure~\ref{fig:qpu_text_case} shows another case, \emph{Le venticinque stanze}: the QPU returns the classical $9$-page MIS exactly, a symmetric set of nine stanze spanning all three floors of the building (\emph{piano terra}, \emph{secondo piano}, \emph{sottotetto}) and three positions across the building's depth (\emph{nord}, \emph{cortile}, \emph{sud}). At register fidelity $1.0$ --- which holds here by construction --- the atom array's blockade pattern reproduces the text graph edge-for-edge, and the adiabatic protocol drives the system into the MIS of \emph{this} text.

Scaling forward, the natural target for outperformance over classical heuristics is not 2D-UDG --- which is classically tractable at any desired accuracy \cite{cazals2025gadgets} --- but bounded-degree non-UDG graphs at high $N$, where no PTAS is known and classical heuristics empirically slow. A random $k$-regular construction at $N \approx 100$--$200$ and $k \approx 10$ is one potential target in that direction; the 256-qubit neutral-atom processor of Leclerc \emph{et al.}\ \cite{leclerc2026magnet256} is the hardware reach. Within 2D-UDG, a $16 \times 16$ extended-king text ($N=256$, $d \approx 0.33$) remains a useful scale-up benchmark --- it shows the hardware can be loaded at the new size --- even if asymptotic outperformance is not in this class.

% ─── Section: Beyond 2D ─────────────────────────────────────────────────

\section{A third dimension: for the atoms and for the texts}\label{sec:beyond2d}

\begin{figure}[!b]
\centering
\includegraphics[width=\textwidth]{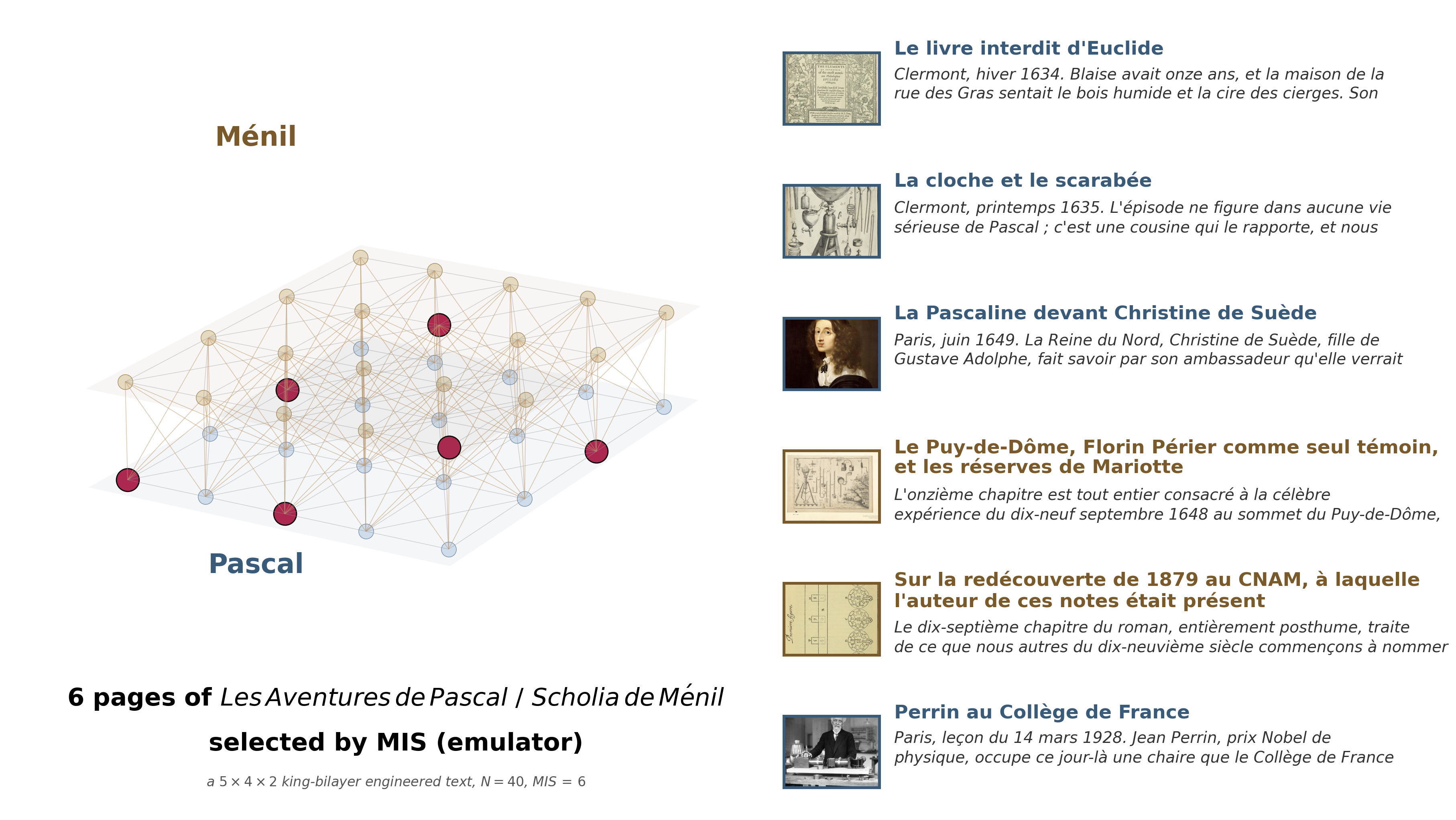}
\caption{\emph{Les Aventures de Pascal} (lower layer, blue) and \emph{Scholia de M\'enil} (upper layer, amber): the $5\!\times\!4\!\times\!2$ king-bilayer engineered text, $N{=}40$, with the six pages of the MIS in red. The right panel lists the MIS pages in deposited reading order. Illustrations, top to bottom: Billingsley's \emph{Euclid}, 1570; Boyle, \emph{New Experiments}, 1660; S. Bourdon, portrait of Queen Christina of Sweden; Deulland, plate of Pascal's Puy-de-D\^ome experiment from B\'elidor's \emph{Architecture hydraulique}; Belair, \emph{Machine arithm\'etique}; J. Perrin, 1927 --- all public-domain, Wikimedia Commons.}
\label{fig:pascal_menil_3d}
\end{figure}

The natural texts and the $k$-NN engineered texts all yield $k$-NN graphs of sentence embeddings, not unit-disk graphs. The SA recall ceiling cited in \S\ref{sec:quantum} is moreover \emph{genre-independent}: across frame-tale (\emph{Decameron}), narrative poem (\emph{Divina Commedia}), encyclopedic miscellany (\emph{Noctes Atticae}), aphorism (\emph{Novum Organum}), and scientific dialogue (Galileo's \emph{Discorsi}), at $N=100$ recall sits in the band $0.21$--$0.27$ at density $d \approx 0.10$--$0.14$; and a random $12$-regular graph at $N{=}75$ (no community structure) gives SA recall $0.21$ in the same band. The ceiling is \emph{geometric} --- a property of bounded-degree graphs whose degree exceeds the 2D-UDG kissing number of $6$, not a literary failure of natural-text clustering --- and no pulse engineering, variational refinement, or noise tuning can recover the edges lost at the register-embedding stage. Dalyac and Henriet \cite{dalyac2022embed3d} prove that any graph of maximum degree $\Delta \leq 6$ admits an exact 3D-array embedding (unlike 2D). Our $k$-NN text graphs typically exceed this sufficient condition --- symmetrised $k{=}8$ neighbours can produce maximum degrees well above $6$ --- so the Dalyac--Henriet theorem does not apply directly. Table~\ref{tab:2d_2L_3d_ladder} below should be read as empirical evidence that the third dimension relaxes the 2D recall ceiling, not as a theorem guaranteeing exact recovery.

A distinctive property of neutral-atom quantum processors --- unique among the current quantum-computing platforms (superconducting qubits, trapped ions, photonics, semiconductor spins) --- is that the register itself can be three-dimensional: optical tweezers position atoms in arbitrary 3D arrangements at the same micron scale that fixes the blockade radius \cite{browaeys2020many}. 3D registers have been demonstrated in the laboratory \cite{barredo2018threedim} but are not yet exposed on commercial cloud QPUs --- FRESNEL itself is 2D, and the discussion in this section is therefore forward-looking, anchored on hardware roadmaps rather than on present submissions.

Before full 3D register reconfigurability becomes available, a simpler intermediate device could already capture part of the 3D benefit: two parallel atom arrays separated by a distance $h \leq R_b$, allowing inter-layer blockade while requiring only a modest extension of existing single-plane trapping optics. To quantify the gain of bilayer and full 3D over the 2D baseline, we ran simulated annealing in three lattice modes --- 2D, bilayer (two stacked planes at $h = 0.80\,R_b$), and 3D --- on the natural-text rows where 2D recall sits well below the $0.5$ threshold. Table~\ref{tab:2d_2L_3d_ladder} reports the resulting edge recall: the lift is monotonic across the dimensional containment $\text{UDG} \subset \text{unit-ball-in-slab} \subset \text{unit-ball-in-}\mathbb{R}^3$ on every row, with bilayer closing roughly $30$--$50\%$ of the 2D-to-3D gap.

\begin{table}[!htbp]
\centering
\small
\begin{tabular}{lcccccc}
\toprule
\textbf{Text} & \textbf{$k$} & \textbf{$N$} & $d$ & \textbf{2D recall} & \textbf{2L recall} & \textbf{3D recall} \\
\midrule
Ausonius \emph{Epigr.} & 8 & 27 & 0.464 & 0.344 & 0.442 & 0.571 \\
Dante \emph{Inferno} & 8 & 34 & 0.330 & 0.368 & 0.465 & 0.546 \\
Augustine \emph{Confessiones} & 8 & 38 & 0.293 & 0.364 & 0.466 & 0.583 \\
Galileo \emph{Dialogo} & 8 & 65 & 0.079 & 0.497 & 0.515 & 0.558 \\
Boethius \emph{Consolatio} & 8 & 72 & 0.146 & 0.339 & 0.392 & 0.543 \\
\bottomrule
\end{tabular}
\caption{Edge-recall ladder for the five natural-text rows where 3D register escape lifts recall above the $0.5$ threshold (the level above which the QPU-recovered backbone substantially matches the canonical text-MIS). Discrete-site simulated-annealing embedder \cite{dalyac2022embed3d} in 2D, bilayer 2L ($h = 0.80\,R_b$), and 3D modes; spring-layout init, $30{,}000$ iterations $\times\,15$ restarts.}
\label{tab:2d_2L_3d_ladder}
\end{table}

The 3D recovery has been validated end-to-end on EMU for three natural texts of the corpus. With the baseline adiabatic pulse, the 3D-embedded Ausonius register ($N=27$, 3D recall $0.571$) recovers $6$ of the $10$ canonical text-MIS pages (canonical ratio $0.60$); the 3D-embedded Augustine register ($N=38$, recall $0.583$) recovers $6$ of the $12$ canonical pages ($0.50$); and the 3D-embedded Boethius register ($N=72$, recall $0.543$) recovers $13$ of the $14$ canonical pages (ratio $0.929$) --- the cleanest of the three, with the low valid-IS rate at this larger graph size reflecting the register's incomplete coverage of canonical edges. The Pascal--M\'enil bilayer (Figure~\ref{fig:pascal_menil_3d}) realises a $5{\times}4{\times}2$ king lattice whose graph is an \emph{exact} bilayer unit-ball graph by construction; run on EMU with the standard adiabatic pulse (intra-layer spacing $a = 5\,\mu$m, layer separation $h = 3\,\mu$m, $R_b = 8\,\mu$m, $1000$ shots; the values $a$ and $h$ here are graph-embedding parameters of a forward-looking bilayer geometry on EMU, not currently exposed FRESNEL loading constraints), it recovers the classical MIS exactly (ratio $1.00$, valid-IS fraction $78.6\%$). Four Pascal pages alternate with two Ménil scholia, the layers commenting on each other at chapter-matched positions. Figure~\ref{fig:pascal_menil_3d} is the controlled case: unlike the three natural-text rows above, where bilayer/3D improves a lossy embedding, the Pascal--M\'enil bilayer is designed directly as an exact bilayer unit-ball text graph, and the canonical MIS is recovered with no embedding loss. Each of these rows crosses the canonical-MIS recovery threshold that 2D embedding could not reach.

3D entanglement grows faster than 2D under the same protocol, so MPS-based emulators that handle $\sim 100$ atoms efficiently in 2D saturate at substantially smaller scales in 3D: a 100-qubit 3D Rydberg machine would likely already escape classical exact simulation, making 3D-capable hardware the natural next target for literary-MIS benchmarks. Staying in 2D and encoding non-UDG graphs through ancilla-gadget constructions \cite{nguyen2023arbitrary,cazals2025gadgets} performs poorly at FRESNEL-scale qubit counts, so we favour the native 3D path. End-to-end QPU validation on bilayer or 3D-capable Rydberg hardware is the natural next experimental step.

% ─── Section: Conclusion ──────────────────────────────────────────────

\section{Envoi}

The category this paper proposed in the Introduction --- a boundary object between neutral-atom MIS and the structural analysis of literary texts --- is now populated densely enough to be tested. Across the realised registers, the processor returned valid independent-set distributions and the sampling outputs (shot histograms, Hamming-weight statistics, register diagnostics) that classical ILP does not produce. Where natural-text recovery is partial, the bottleneck sits at the 2D embedding stage, not in the hardware; the forward-looking extension reported in \S\ref{sec:beyond2d} --- 3D register geometries that lift canonical-backbone recovery on the emulator, with a bilayer construction recovering its backbone exactly --- targets the hardware generation in which the planar embedding ceiling identified here no longer bounds the campaign.

Two openings are offered beyond this paper. Rigidity $\rho$, the (text, embedding)-pair metric introduced here, is portable: any deposited corpus with a fixed embedder and an ILP-verified MIS can be characterised the same way, and we hope the field tests it against kinds of textual organisation we did not run. The same framework also crosses out of text. A pilot extension in preparation applies the MIS-backbone construction to a visual dossier: artworks as nodes, multimodal visual embeddings layered with the ICON~2.0 ontology \cite{sartini2023icon} and IICONGRAPH constructions \cite{sartini2024iicongraph,baroncini2025monastic}, MIS as the structural backbone of an iconographic canon --- to our knowledge a first. Beyond these openings, we invite the digital-humanities, quantum-hardware, and benchmark communities to extend the corpus to harder canonical instances and to track its behaviour across processor generations as the hardware scales.

A note on scale, for the humanist reader. In the quantum-computing community the central scaling challenge is to grow from today's $\sim$100-atom machines to thousands and ultimately hundreds of thousands of atoms or qubits, and neutral atoms held in optical tweezers are uniquely positioned to reach that regime: defect-free arrays of $1024$ atoms in a cryogenic environment \cite{lim2026thousand} and 6100-atom alkaline-earth tweezer arrays \cite{manetsch2025tweezer} are already experimentally demonstrated, and are expected within the next few years to reach commercial cloud-accessible platforms of the kind used in this paper. As the platform grows, the corpus we deposit grows with it --- larger registers admit longer natural texts at finer chunking, denser engineered graphs, and constraints whose canonical sizes were too large to instantiate at $N=100$.

Each generation reads its inherited corpus through the tools it has; with cloud-exposed neutral-atom processors and agentic LLM interfaces, that toolset has just acquired a new member. Outstanding instruments are now within reach of literary scholars, historians, musicologists, and philologists alike --- from here it is for the imagination to take the lead, and we know that imagination has no limits.

% ─── Data Availability ───────────────────────────────────────────────────

\vspace{1em}
\section{Data availability}\label{sec:data}

The QOuLiPo corpus is openly deposited on Zenodo \cite{jurczak2026qoulipo_dataset} (DOI \href{https://doi.org/10.5281/zenodo.20074378}{10.5281/zenodo.20074378}, CC-BY-4.0): 29 engineered QOuLiPo folders, 11 natural-text companion corpora, per-folder \texttt{metadata.json}, explicit-edge-list graph files, ILP-verified MIS solutions and embedding distances (\texttt{mis\_analysis\_all.csv}), a strict consistency verifier (\texttt{verify.py --strict --lint}), and a standalone single-page HTML navigator. The OCR-derived digitisation of the 1544 Giambullari is at \cite{jurczak2026lightonocr}.

\vspace{0.5em}
\section{Disclosures}

The author is a co-founder of Pasqal, the company whose FRESNEL\_CAN1 neutral-atom processor and EMU\_MPS cloud emulator are used throughout this paper, and a co-founder and managing partner at Quantonation, a deep-tech venture capital firm with positions in quantum hardware companies including Pasqal. Both relationships are disclosed for transparency; all QPU results and the bulk of the EMU runs were obtained on standard Pasqal Cloud accounts available to any external user, with a subset of larger EMU instances run on commercial H100/H200 cloud resources (see \S\ref{sec:pipeline}). The 1544 \emph{editio princeps} of Giambullari's \emph{Del sito, forma, \& misure dello Inferno di Dante} and the 1517 Aldine \emph{Opera} of Ausonius are part of the author's personal collection of Renaissance and humanist books; these physical witnesses were the originals photographed and OCR'd for the corpus described in \S2 and Appendix~\ref{app:corpus}.

\paragraph{Use of generative AI.} The QOuLiPo prose deposited with this paper was generated by Anthropic's Claude (Sonnet 4.5 / Opus 4.6 / Opus 4.7) under the author's design, validation, and editorial guidance, as described in \S\ref{sec:qoulipo}. All FRESNEL\_CAN1 and EMU\_MPS submissions reported in \S\ref{sec:quantum} and \S\ref{sec:beyond2d} were authored through an agentic Claude Opus loop calling Pulser and the Pasqal Cloud, as described in \S\ref{sec:intro}.
\vspace{0.5em}
\section{Acknowledgments}\label{sec:acknowledgments}

The author thanks Constantin Dalyac, Lo\"{\i}c Henriet, Alexandre Dauphin, and Mourad Beji at Pasqal for technical discussions on FRESNEL, register embedding, and noise characterisation; Igor Carron at LightOn for discussions on the VLM-OCR pipeline used to digitise the Giambullari and Ausonius sources; and Bernard Mar\'echal (\href{https://www.zazipo.net/+-Bernard-J-Marechal-+}{zazipo.net}) for inspiring the constrained-writing framing.% ─── References ─────────────────────────────────────────────────────────

\bibliographystyle{ieeetr}
\bibliography{references}

@book{roubaud1967epsilon,
  author    = {Roubaud, Jacques},
  title     = {$\in$},
  publisher = {Gallimard},
  address   = {Paris},
  year      = {1967},
}

@article{feynman1982simulating,
  author  = {Feynman, Richard P.},
  title   = {Simulating Physics with Computers},
  journal = {International Journal of Theoretical Physics},
  volume  = {21},
  number  = {6--7},
  pages   = {467--488},
  year    = {1982},
  doi     = {10.1007/BF02650179},
}

@article{schollwock2011dmrg,
  author  = {Schollw\"{o}ck, Ulrich},
  title   = {The Density-Matrix Renormalization Group in the Age of Matrix Product States},
  journal = {Annals of Physics},
  volume  = {326},
  number  = {1},
  pages   = {96--192},
  year    = {2011},
  doi     = {10.1016/j.aop.2010.09.012},
  url     = {https://doi.org/10.1016/j.aop.2010.09.012},
}

@misc{pasqal2024emumps,
  author       = {{Pasqal}},
  title        = {{emu-mps}: A Matrix-Product-State Emulator for Neutral-Atom Quantum Processors},
  year         = {2024},
  howpublished = {\url{https://github.com/pasqal-io/emulators}},
  note         = {Open-source tensor-network backend for Pasqal Cloud},
}

@article{nemhauser1975vertexpack,
  author  = {Nemhauser, George L. and Trotter, Leslie E., Jr.},
  title   = {Vertex Packings: Structural Properties and Algorithms},
  journal = {Mathematical Programming},
  volume  = {8},
  number  = {1},
  pages   = {232--248},
  year    = {1975},
  doi     = {10.1007/BF01580444},
}

@article{hammer1984persistency,
  author  = {Hammer, Peter L. and Hansen, Pierre and Simeone, Bruno},
  title   = {Roof Duality, Complementation and Persistency in Quadratic 0--1 Optimization},
  journal = {Mathematical Programming},
  volume  = {28},
  number  = {2},
  pages   = {121--155},
  year    = {1984},
  doi     = {10.1007/BF02612354},
}

@article{boros2002pbo,
  author  = {Boros, Endre and Hammer, Peter L.},
  title   = {Pseudo-{B}oolean Optimization},
  journal = {Discrete Applied Mathematics},
  volume  = {123},
  number  = {1--3},
  pages   = {155--225},
  year    = {2002},
  doi     = {10.1016/S0166-218X(01)00341-9},
}

@article{blei2003lda,
  author  = {Blei, David M. and Ng, Andrew Y. and Jordan, Michael I.},
  title   = {Latent {D}irichlet Allocation},
  journal = {Journal of Machine Learning Research},
  volume  = {3},
  pages   = {993--1022},
  year    = {2003},
  url     = {https://jmlr.org/papers/v3/blei03a.html},
}

@article{cazals2025mis,
  author  = {Cazals, Pierre and François, Aymeric and Henriet, Loïc and Leclerc, Lucas and Marin, Malory and Naghmouchi, Yassine and da Silva Coelho, Wesley and Sikora, Florian and Vitale, Vittorio and Watrigant, Rémi and Witt Garzillo, Monique and Dalyac, Constantin},
  title   = {Identifying Hard Native Instances for the Maximum-Independent-Set Problem on Neutral-Atom Quantum Processors},
  journal = {Physical Review Applied},
  volume  = {25},
  number  = {3},
  year    = {2026},
  doi     = {10.1103/8gjv-ll1f},
  url     = {https://doi.org/10.1103/8gjv-ll1f},
}

@article{shiraishi2026mcp,
  author  = {Shiraishi, Masaki and Hamamura, Ikko and Ishigaki, Tatsuya and Kadowaki, Tadashi},
  title   = {A Model Context Protocol Server for Quantum Execution in Hybrid Quantum-HPC Environments},
  journal = {arXiv preprint arXiv:2604.08318},
  year    = {2026},
  url     = {https://arxiv.org/abs/2604.08318},
}

@article{leclerc2026magnet256,
  author  = {Leclerc, Lucas and Julià-Farré, Sergi and Silva Freitas, Gabriel and Villaret, Guillaume and Albrecht, Boris and Béguin, Lucas and others},
  title   = {One-to-One Quantum Simulation of the Low-Dimensional Frustrated Quantum Magnet {TmMgGaO}$_4$ with 256 Qubits},
  journal = {arXiv preprint arXiv:2603.20372},
  year    = {2026},
  url     = {https://arxiv.org/abs/2603.20372},
}

@article{silverio2022pulser,
  author  = {Silvério, Henrique and Grijalva, Sebastián and Dalyac, Constantin and Leclerc, Lucas and Karalekas, Peter J. and Shammah, Nathan and Beji, Mourad and Henry, Louis-Paul and Henriet, Loïc},
  title   = {Pulser: An Open-Source Package for the Design of Pulse Sequences in Programmable Neutral-Atom Arrays},
  journal = {Quantum},
  volume  = {6},
  pages   = {629},
  year    = {2022},
  url     = {https://doi.org/10.22331/q-2022-01-24-629},
}

@article{henriet2020quantum,
  author  = {Henriet, Loïc and Beguin, Lucas and Signoles, Adrien and Lahaye, Thierry and Browaeys, Antoine and Reymond, Georges-Olivier and Jurczak, Christophe},
  title   = {Quantum Computing with Neutral Atoms},
  journal = {Quantum},
  volume  = {4},
  pages   = {327},
  year    = {2020},
  url     = {https://doi.org/10.22331/q-2020-09-21-327},
}

@article{pichler2018quantum,
  author  = {Pichler, Hannes and Wang, Sheng-Tao and Zhou, Leo and Choi, Soonwon and Lukin, Mikhail D.},
  title   = {Quantum Optimization for Maximum Independent Set Using {R}ydberg Atom Arrays},
  journal = {arXiv preprint arXiv:1808.10816},
  year    = {2018},
  url     = {https://arxiv.org/abs/1808.10816},
}

@article{ebadi2022quantum,
  author  = {Ebadi, Sepehr and Keesling, Alexander and Cain, Madelyn and Wang, Tout T. and Levine, Harry and Bluvstein, Dolev and Semeghini, Giulia and Omran, Ahmed and Liu, Jin-Guo and Samajdar, Rhine and Luo, Xiu-Zhe and Nash, Beatrice and Gao, Xun and Barak, Boaz and Farhi, Edward and Sachdev, Subir and Gemelke, Nathan and Zhou, Leo and Choi, Soonwon and Pichler, Hannes and Wang, Sheng-Tao and Greiner, Markus and Vuletić, Vladan and Lukin, Mikhail D.},
  title   = {Quantum Optimization of Maximum Independent Set Using {R}ydberg Atom Arrays},
  journal = {Science},
  volume  = {376},
  number  = {6598},
  pages   = {1209--1215},
  year    = {2022},
  url     = {https://doi.org/10.1126/science.abo6587},
}

@book{rockwell2016hermeneutica,
  author    = {Rockwell, Geoffrey and Sinclair, Stéfan},
  title     = {Hermeneutica: Digital Humanities, Interpretation and Technology},
  publisher = {MIT Press},
  year      = {2016},
}

@book{moretti2013distant,
  author    = {Moretti, Franco},
  title     = {Distant Reading},
  publisher = {Verso},
  year      = {2013},
}

@book{jockers2013macroanalysis,
  author    = {Jockers, Matthew L.},
  title     = {Macroanalysis: Digital Methods and Literary History},
  publisher = {University of Illinois Press},
  year      = {2013},
}

@book{underwood2019distant,
  author    = {Underwood, Ted},
  title     = {Distant Horizons: Digital Evidence and Literary Change},
  publisher = {University of Chicago Press},
  year      = {2019},
}

@article{uckan2020extractive,
  author  = {Uçkan, Tuba and Karcı, Ali},
  title   = {Extractive Multi-Document Text Summarization Based on Graph Independent Sets},
  journal = {Egyptian Informatics Journal},
  volume  = {21},
  number  = {3},
  pages   = {145--157},
  year    = {2020},
  url     = {https://doi.org/10.1016/j.eij.2019.12.002},
}

@article{hark2025mis,
  author  = {Hark, Cihan},
  title   = {Using Graph-Based Maximum Independent Sets with Large Language Models for Extractive Text Summarization},
  journal = {Applied Sciences},
  volume  = {15},
  number  = {12},
  pages   = {6395},
  year    = {2025},
  url     = {https://doi.org/10.3390/app15126395},
}

@book{berkman2022oulipo,
  author    = {Berkman, Natalie},
  title     = {{OuLiPo} and the Mathematics of Literature},
  publisher = {Peter Lang},
  year      = {2022},
}

@book{queneau1961sonnets,
  author    = {Queneau, Raymond},
  title     = {Cent Mille Milliards de Poèmes},
  publisher = {Gallimard},
  year      = {1961},
}

@book{perec1969disparition,
  author    = {Perec, Georges},
  title     = {La Disparition},
  publisher = {Denoël},
  year      = {1969},
}

@book{perec1978vie,
  author    = {Perec, Georges},
  title     = {La Vie mode d'emploi},
  publisher = {Hachette},
  address   = {Paris},
  year      = {1978},
}

@book{calvino1973castello,
  author    = {Calvino, Italo},
  title     = {Il castello dei destini incrociati},
  publisher = {Einaudi},
  address   = {Torino},
  year      = {1973},
}

@book{calvino1979notte,
  author    = {Calvino, Italo},
  title     = {Se una notte d'inverno un viaggiatore},
  publisher = {Einaudi},
  address   = {Torino},
  year      = {1979},
}

@book{calvino1988lezioni,
  author    = {Calvino, Italo},
  title     = {Lezioni americane: sei proposte per il prossimo millennio},
  publisher = {Garzanti},
  address   = {Milano},
  year      = {1988},
}

@article{wang2024multilinguale5instruct,
  title        = {Multilingual {E5} Text Embeddings: A Technical Report},
  author       = {Wang, Liang and Yang, Nan and Huang, Xiaolong and Yang, Linjun and Majumder, Rangan and Wei, Furu},
  journal      = {arXiv preprint arXiv:2402.05672},
  year         = {2024},
}

@article{chen2024bgem3,
  title        = {{BGE M3-Embedding}: Multi-Lingual, Multi-Functionality, Multi-Granularity Text Embeddings Through Self-Knowledge Distillation},
  author       = {Chen, Jianlv and Xiao, Shitao and Zhang, Peitian and Luo, Kun and Lian, Defu and Liu, Zheng},
  journal      = {arXiv preprint arXiv:2402.03216},
  year         = {2024},
}

@misc{yu2024arcticembed,
  title        = {{Arctic-Embed 2.0}: Multilingual Retrieval Without Compromise},
  author       = {Yu, Puxuan and Merrick, Luke and Nuti, Gaurav and Campos, Daniel},
  year         = {2024},
  publisher    = {Snowflake AI Research},
  url          = {https://huggingface.co/Snowflake/snowflake-arctic-embed-l-v2.0},
}

@inproceedings{enevoldsen2025mmteb,
  title        = {{MMTEB}: Massive Multilingual Text Embedding Benchmark},
  author       = {Enevoldsen, Kenneth and others},
  booktitle    = {International Conference on Learning Representations (ICLR)},
  year         = {2025},
  url          = {https://openreview.net/forum?id=zl3pfz4VCV},
}

@misc{jurczak2026lightonocr,
  author       = {Jurczak, Christophe},
  title        = {A 2.48\% {CER} {VLM}-{OCR} pipeline for a heavily accented 1544 {F}lorentine treatise},
  year         = {2026},
  month        = apr,
  publisher    = {Humanities Commons},
  doi          = {10.17613/dfax8-7vd97},
  url          = {https://zenodo.org/records/19503091},
  note         = {Preprint. \url{https://zenodo.org/records/19503091}},
}

@misc{jurczak2026qoulipo_dataset,
  author    = {Jurczak, Christophe},
  title     = {{QOuLiPo} Corpus --- Canonical Text-Graph Data},
  year      = {2026},
  publisher = {Zenodo},
  version   = {v2.0},
  doi       = {10.5281/zenodo.20074378},
  url       = {https://doi.org/10.5281/zenodo.20074378},
  note      = {Dataset. \url{https://doi.org/10.5281/zenodo.20074378}},
}

@incollection{levyleblond2022dante,
  author       = {L\'evy-Leblond, Jean-Marc},
  title        = {A Mathematical Physicist in Hell: {G}alileo on the Geometry of {D}ante's {I}nferno},
  booktitle    = {Imagine Math 8: Dreaming Venice},
  editor       = {Emmer, Michele and Abate, Marco},
  publisher    = {Springer},
  year         = {2022},
  doi          = {10.1007/978-3-030-92690-8\_29},
  url          = {https://doi.org/10.1007/978-3-030-92690-8_29},
}

@book{peterson2011galileo,
  author       = {Peterson, Mark A.},
  title        = {Galileo's Muse: {R}enaissance Mathematics and the Arts},
  publisher    = {Harvard University Press},
  year         = {2011},
  isbn         = {9780674059726},
  url          = {https://www.hup.harvard.edu/books/9780674059726},
}

@book{netz2011palimpsest,
  author       = {Netz, Reviel and Noel, William},
  title        = {The {A}rchimedes Codex: Revealing the Secrets of the World's Greatest Palimpsest},
  publisher    = {Da Capo Press},
  year         = {2011},
}

@misc{ausonius_cento_preface,
  author       = {Ausonius, Decimus Magnus},
  title        = {Praefatio ad {C}entonem {N}uptialem},
  year         = {ca.~370},
  note         = {Modern critical edition: R.~P.~H.~Green (ed.), \emph{The Works of Ausonius}, Oxford: Clarendon Press, 1991, pp.~132--134; Loeb edition: H.~G.~Evelyn-White (trans.), \emph{Ausonius}, vol.~1, Cambridge MA: Harvard University Press, 1919/1985, pp.~370--373. Photographed in this work from the Aldine \emph{editio princeps}, Venice, 1517.},
}

@phdthesis{blossier2009oulipiens,
  author       = {Blossier-Jacquemot, Anne},
  title        = {Les {O}ulipiens antiques: pour une anthropologie des pratiques d'\'ecriture \`a contraintes dans l'antiquit\'e},
  school       = {Universit\'e {P}aris {VII} {D}iderot},
  year         = {2009},
  url          = {https://www.theses.fr/2009PA070056},
}

@article{dalyac2022embed3d,
  author       = {Dalyac, Constantin and Henriet, Lo\"ic},
  title        = {Embedding the {MIS} problem for non-local graphs with bounded degree using {3D} arrays of atoms},
  journal      = {arXiv preprint arXiv:2209.05164},
  year         = {2022},
}

@article{dalyac2024graph,
  author       = {Dalyac, Constantin and Leclerc, Lucas and Vignoli, Louis and Djellabi, Mehdi and da Silva Coelho, Wesley and Ximenez, Bruno and Dareau, Alexandre and Dreon, Davide and Elfving, Vincent E. and Signoles, Adrien and Henry, Louis-Paul and Henriet, Lo\"ic},
  title        = {Graph Algorithms with Neutral Atom Quantum Processors},
  journal      = {European Physical Journal A},
  volume       = {60},
  pages        = {177},
  year         = {2024},
  doi          = {10.1140/epja/s10050-024-01385-5},
  eprint       = {2403.11931},
  archivePrefix = {arXiv},
  url          = {https://doi.org/10.1140/epja/s10050-024-01385-5},
}

@article{nguyen2023arbitrary,
  author       = {Nguyen, Minh-Thi and Liu, Jin-Guo and Wurtz, Jonathan and Lukin, Mikhail D. and Wang, Sheng-Tao and Pichler, Hannes},
  title        = {Quantum optimization with arbitrary connectivity using {R}ydberg atom arrays},
  journal      = {PRX Quantum},
  volume       = {4},
  number       = {1},
  pages        = {010316},
  year         = {2023},
  doi          = {10.1103/PRXQuantum.4.010316},
}

@article{gonzalez1985kcenter,
  author = {Gonzalez, Teofilo F.},
  title = {Clustering to minimize the maximum intercluster distance},
  journal = {Theoretical Computer Science},
  volume = {38},
  pages = {293--306},
  year = {1985},
  doi = {10.1016/0304-3975(85)90224-5},
}

@article{hochbaum1985kcenter,
  author = {Hochbaum, Dorit S. and Shmoys, David B.},
  title = {A best possible heuristic for the $k$-center problem},
  journal = {Mathematics of Operations Research},
  volume = {10},
  number = {2},
  pages = {180--184},
  year = {1985},
  doi = {10.1287/moor.10.2.180},
}

@article{nemhauser1978submodular,
  author = {Nemhauser, George L. and Wolsey, Laurence A. and Fisher, Marshall L.},
  title = {An analysis of approximations for maximizing submodular set functions -- {I}},
  journal = {Mathematical Programming},
  volume = {14},
  pages = {265--294},
  year = {1978},
  doi = {10.1007/BF01588971},
}

@article{krause2014submodular,
  author = {Krause, Andreas and Golovin, Daniel},
  title = {Submodular function maximization},
  journal = {Tractability: Practical Approaches to Hard Problems},
  pages = {71--104},
  year = {2014},
  publisher = {Cambridge University Press},
}

@article{jaksch2000rydberg,
  author = {Jaksch, D. and Cirac, J. I. and Zoller, P. and Rolston, S. L. and C\^ot\'e, R. and Lukin, M. D.},
  title = {Fast quantum gates for neutral atoms},
  journal = {Phys. Rev. Lett.},
  volume = {85},
  number = {10},
  pages = {2208--2211},
  year = {2000},
  doi = {10.1103/PhysRevLett.85.2208},
}

@article{urban2009observation,
  author = {Urban, E. and Johnson, T. A. and Henage, T. and Isenhower, L. and Yavuz, D. D. and Walker, T. G. and Saffman, M.},
  title = {Observation of {R}ydberg blockade between two atoms},
  journal = {Nature Physics},
  volume = {5},
  pages = {110--114},
  year = {2009},
  doi = {10.1038/nphys1178},
}

@article{browaeys2020many,
  author = {Browaeys, Antoine and Lahaye, Thierry},
  title = {Many-body physics with individually controlled {R}ydberg atoms},
  journal = {Nature Physics},
  volume = {16},
  pages = {132--142},
  year = {2020},
  doi = {10.1038/s41567-019-0733-z},
}

@article{amancio2015probing,
  author  = {Amancio, Diego R.},
  title   = {Probing the topological properties of complex networks modeling short written texts},
  journal = {PLoS ONE},
  volume  = {10},
  number  = {2},
  pages   = {e0118394},
  year    = {2015},
  doi     = {10.1371/journal.pone.0118394},
}

@article{amancio2013network,
  author  = {Amancio, Diego R. and Oliveira Jr., Osvaldo N. and Costa, Luciano da F.},
  title   = {Identification of literary movements using complex networks to represent texts},
  journal = {New Journal of Physics},
  volume  = {15},
  pages   = {043048},
  year    = {2013},
  doi     = {10.1088/1367-2630/15/4/043048},
}

@inproceedings{elson2010extracting,
  author    = {Elson, David K. and Dames, Nicholas and McKeown, Kathleen R.},
  title     = {Extracting social networks from literary fiction},
  booktitle = {Proceedings of the 48th Annual Meeting of the ACL},
  pages     = {138--147},
  year      = {2010},
}

@article{labatut2019extraction,
  author  = {Labatut, Vincent and Bost, Xavier},
  title   = {Extraction and analysis of fictional character networks: a survey},
  journal = {ACM Computing Surveys},
  volume  = {52},
  number  = {5},
  pages   = {89},
  year    = {2019},
  doi     = {10.1145/3344548},
}

@article{antiqueira2009complex,
  author  = {Antiqueira, Lucas and Nunes, Maria das Gra{\c{c}}as V. and Oliveira Jr., Osvaldo N. and Costa, Luciano da F.},
  title   = {Strong correlations between text quality and complex networks features},
  journal = {Physica A: Statistical Mechanics and its Applications},
  volume  = {388},
  number  = {12},
  pages   = {2603--2611},
  year    = {2009},
  doi     = {10.1016/j.physa.2009.02.048},
}

@article{moretti2011network,
  author  = {Moretti, Franco},
  title   = {Network theory, plot analysis},
  journal = {New Left Review},
  volume  = {68},
  pages   = {80--102},
  year    = {2011},
}

@article{meichanetzidis2023grammar,
  author  = {Meichanetzidis, Konstantinos and Toumi, Alexis and de Felice, Giovanni and Coecke, Bob},
  title   = {Grammar-aware sentence classification on quantum computers},
  journal = {Quantum Machine Intelligence},
  volume  = {5},
  number  = {1},
  pages   = {16},
  year    = {2023},
  note    = {arXiv:2012.03756},
  doi     = {10.1007/s42484-023-00097-1},
  url     = {https://doi.org/10.1007/s42484-023-00097-1},
}

@book{oulipo1981atlas,
  author    = {{OuLiPo}},
  title     = {Atlas de litt\'erature potentielle},
  publisher = {Gallimard},
  address   = {Paris},
  year      = {1981},
}

@article{barzen2019quantum,
  author  = {Barzen, Johanna and Leymann, Frank},
  title   = {Quantum humanities: a vision for quantum computing in digital humanities},
  journal = {SICS Software-Intensive Cyber-Physical Systems},
  volume  = {35},
  pages   = {153--158},
  year    = {2019},
  doi     = {10.1007/s00450-019-00419-4},
}

@incollection{barzen2021digital,
  author    = {Barzen, Johanna},
  title     = {From Digital Humanities to Quantum Humanities: Potentials and Applications},
  booktitle = {Quantum Computing in the Arts and Humanities},
  editor    = {Miranda, Eduardo R.},
  publisher = {Springer},
  year      = {2022},
  note      = {arXiv:2103.11825},
}

@book{miranda2022qc_arts_humanities,
  editor    = {Miranda, Eduardo R.},
  title     = {Quantum Computing in the Arts and Humanities: An Introduction to Core Concepts, Theory and Applications},
  publisher = {Springer},
  year      = {2022},
  doi       = {10.1007/978-3-030-95538-0},
}

@article{lorenz2023qnlp,
  author  = {Lorenz, Robin and Pearson, Anna and Meichanetzidis, Konstantinos and Kartsaklis, Dimitri and Coecke, Bob},
  title   = {{QNLP} in Practice: Running Compositional Models of Meaning on a Quantum Computer},
  journal = {Journal of Artificial Intelligence Research},
  volume  = {76},
  pages   = {1305--1342},
  year    = {2023},
  note    = {arXiv:2102.12846},
}

@article{wazni2024pronoun,
  author  = {Wazni, Hadi and Lo, Kin Ian and McPheat, Lachlan and Sadrzadeh, Mehrnoosh},
  title   = {Large-scale structure-aware pronoun resolution using quantum natural language processing},
  journal = {Quantum Machine Intelligence},
  volume  = {6},
  number  = {2},
  pages   = {60},
  year    = {2024},
  doi     = {10.1007/s42484-024-00193-w},
  url     = {https://doi.org/10.1007/s42484-024-00193-w},
}

@inproceedings{sartini2024iicongraph,
  author    = {Sartini, Bruno},
  title     = {{IICONGRAPH}: improved Iconographic and Iconological Statements in Knowledge Graphs},
  booktitle = {The Semantic Web --- 21st European Semantic Web Conference, {ESWC} 2024, Hersonissos, Crete, Greece, May 26--30, 2024, Proceedings},
  publisher = {Springer},
  year      = {2024},
  note      = {arXiv:2402.00048; dataset Zenodo DOI 10.5281/zenodo.10294589, CC-BY 4.0},
  url       = {https://arxiv.org/abs/2402.00048},
}

@article{sartini2023icon,
  author  = {Sartini, Bruno and Baroncini, Sofia and van Erp, Marieke and Tomasi, Francesca and Gangemi, Aldo},
  title   = {{ICON}: An Ontology for Comprehensive Artistic Interpretations},
  journal = {Journal on Computing and Cultural Heritage},
  volume  = {16},
  number  = {3},
  pages   = {59--76},
  year    = {2023},
  doi     = {10.1145/3594724},
  url     = {https://doi.org/10.1145/3594724},
  note    = {ICON ontology v2.0 at \url{https://w3id.org/icon/ontology}; formalises Panofsky's tripartite pre-iconographic / iconographic / iconological framework},
}

@inproceedings{baroncini2025monastic,
  author    = {Baroncini, Sofia and Mele, Francesco},
  title     = {Describing Monastic Iconography Using Semantic Data: A Preliminary Investigation},
  booktitle = {AIUCD 2025: Conference of the Italian Association for Digital Humanities},
  year      = {2025},
  url       = {https://aiucd2025.dlls.univr.it/assets/pdf/papers/76.pdf},
}

@misc{loeb,
  author       = {{Harvard University Press}},
  title        = {Loeb Classical Library},
  howpublished = {\url{https://www.loebclassics.com/}},
  note         = {Subscription bilingual Greek/Latin–English series; founded 1911},
}

@misc{latinlibrary,
  author       = {{The Latin Library}},
  title        = {The Latin Library: A Free Resource for Latin Texts},
  howpublished = {\url{https://www.thelatinlibrary.com/}},
  note         = {Open-access Latin transcriptions, classical and post-classical},
}

@misc{digiliblt,
  author       = {{DigiLiBLT}},
  title        = {{DigiLiBLT}: Biblioteca Digitale di Testi Latini Tardoantichi},
  howpublished = {\url{https://digiliblt.uniupo.it/}},
  note         = {Universit\`a degli Studi del Piemonte Orientale; open-access digital library of Late Antique Latin texts},
}

@article{vovrosh2025crossover,
  author  = {Vovrosh, Joseph and Mendes-Santos, Tiago and Mamann, Antoine and Bidzhiev, Kemal and Hayes, Matthew and Ximenez, Bruno and B\'eguin, Lucas and Dalyac, Constantin and Dauphin, Alexandre},
  title   = {Resource Assessment of Classical and Quantum Hardware for Post-Quench Dynamics},
  journal = {arXiv preprint arXiv:2511.20388},
  year    = {2025},
  url     = {https://arxiv.org/abs/2511.20388},
}

@article{cazals2025gadgets,
  author        = {Cazals, Pierre and Sorondo, Amalia and Onofre, Victor and Dalyac, Constantin and Coelho, Wesley and Vitale, Vittorio},
  title         = {Quantum Optimization on Rydberg Atom Arrays with Arbitrary Connectivity: Gadgets Limitations and a Heuristic Approach},
  journal       = {Physical Review A},
  volume        = {112},
  pages         = {062416},
  year          = {2025},
  eprint        = {2508.06130},
  archivePrefix = {arXiv},
  primaryClass  = {quant-ph}
}

@book{manning2008ir,
  author    = {Manning, Christopher D. and Raghavan, Prabhakar and Sch\"{u}tze, Hinrich},
  title     = {Introduction to Information Retrieval},
  publisher = {Cambridge University Press},
  year      = {2008},
  address   = {Cambridge}
}

@article{barredo2018threedim,
  author  = {Barredo, Daniel and Lienhard, Vincent and de L\'es\'eleuc, Sylvain and Lahaye, Thierry and Browaeys, Antoine},
  title   = {Synthetic three-dimensional atomic structures assembled atom by atom},
  journal = {Nature},
  volume  = {561},
  number  = {7721},
  pages   = {79--82},
  year    = {2018},
  doi     = {10.1038/s41586-018-0450-2}
}

@article{kommers2025hermeneutics,
  author  = {Kommers, Cody and Ahnert, Ruth and Antoniak, Maria and Benetos, Emmanouil and Benford, Steve and Bunz, Mercedes and Caramiaux, Baptiste and Concannon, Shauna and Disley, Martin and Dobson, James and Du, Yali and Du\'e\~nez-Guzm\'an, Edgar and Francksen, Kerry and Gius, Evelyn and Gray, Jonathan and Heuser, Ryan and Immel, Sarah and So, Richard Jean and Leigh, Sang and Livingston, Dalaki and Long, Hoyt and Martin, Meredith and Meyer, Georgia and Mihai, Daniela and Noel-Hirst, Ashley and Ostherr, Kirsten and Parker, Deven and Qin, Yipeng and Ratcliff, Jessica and Robinson, Emily and Rodriguez, Karina and Sobey, Adam and Underwood, Ted and Vashistha, Aditya and Wilkens, Matthew and Wu, Youyou and Yuan, Zheng and Hemment, Drew},
  title   = {Computational Hermeneutics: Evaluating Generative {AI} as a Cultural Technology},
  journal = {SSRN preprint},
  year    = {2025},
  note    = {SSRN abstract ID 5409144, submitted August 2025},
  url     = {https://papers.ssrn.com/sol3/papers.cfm?abstract_id=5409144}
}

@book{borges1941jardin,
  author    = {Borges, Jorge Luis},
  title     = {El jard\'in de senderos que se bifurcan},
  publisher = {Editorial Sur},
  address   = {Buenos Aires},
  year      = {1941},
  note      = {Collection containing \emph{La biblioteca de Babel} and the title story \emph{El jard\'in de senderos que se bifurcan}; both later included in \emph{Ficciones} (1944).}
}

@article{lim2026thousand,
  author    = {Lim, Desiree and Mamann, Hadriel and Pichard, Gr\'egoire and others},
  title     = {Defect-free arrays at the thousand-atom scale in a 4-{K} cryogenic environment},
  journal   = {arXiv preprint arXiv:2604.07205},
  year      = {2026}
}

@article{manetsch2025tweezer,
  author    = {Manetsch, Hannah J. and Nomura, Gyohei and Bataille, Elie and Leung, Kon H. and Lv, Xudong and Endres, Manuel},
  title     = {A tweezer array with 6,100 highly coherent atomic qubits},
  journal   = {Nature},
  volume    = {647},
  pages     = {60--67},
  year      = {2025},
  doi       = {10.1038/s41586-025-09641-4}
}

% ═══════════════════════════════════════════════════════════════════════════
%  APPENDICES
% ═══════════════════════════════════════════════════════════════════════════

\clearpage

\begin{center}
  {\Large\bfseries Supporting Information}

  \vspace{0.5em}
  \noindent\rule{0.6\textwidth}{0.5pt}

  \vspace{2em}
\end{center}

\noindent The supporting information collects the corpus and methodology references for the main text:

\vspace{1em}
\begin{description}
  \setlength{\itemsep}{0.5em}
  \item[\textbf{A. Corpus.}] The eight natural texts used throughout, with author, date, language, segmentation, description and source edition.
  \item[\textbf{B. Embedding validation.}] Cross-embedder sensitivity study on four natural texts $\times$ three multilingual transformers.
  \item[\textbf{C. \emph{Twelve Hours of a Coastal Town}: full text.}] Complete English mini-text.
  \item[\textbf{D. Classical-benchmark results.}] Classical MIS solver timings (ILP, greedy, simulated annealing).
  \item[\textbf{E. QOuLiPo: full corpus inventory.}] Per-text catalogue of the twenty-nine constrained QOuLiPo texts.
  \item[\textbf{F. Register and pulse optimisation.}] Register engineering (SA, blockade margin, case studies) and pulse design (baseline, variational sweep).
\end{description}

\appendix
\renewcommand{\thesection}{\Alph{section}}

% ─── Appendix A: Corpus ──────────────────────────────────────────────────

\clearpage
\begin{landscape}
\section{Corpus}\label{app:corpus}

\begin{table}[H]
\centering
\footnotesize
\setlength{\tabcolsep}{6pt}
\begin{tabular}{>{\raggedright\arraybackslash}p{3.4cm}>{\raggedright\arraybackslash}p{2.5cm}lcc>{\raggedright\arraybackslash}p{2.4cm}>{\raggedright\arraybackslash}p{6.8cm}>{\raggedright\arraybackslash}p{3.6cm}@{}}
\toprule
\textbf{Text} & \textbf{Author} & \textbf{Date} & \textbf{Lang} & $N$ & \textbf{Unit} & \textbf{What it is} & \textbf{Source / edition} \\
\midrule
\emph{De'l sito, forma, \& misure dello Inferno di Dante} & P.F.~Giambullari & 1544 & IT & 151\,/\,65 & printed page (octavo) & Renaissance Florentine treatise reconstructing the geometry of Dante's Hell from textual clues in the \emph{Divina Commedia}; central document of this edition project. & Neri Dortelata, Firenze, 1544; OCR \cite{jurczak2026lightonocr}. \\[2pt]
\emph{Dialogo sopra i due massimi sistemi del mondo} & Galileo Galilei & 1632 & IT & 65 & 400-word window & Defence of heliocentrism as a four-day conversation between three speakers; the book that triggered Galileo's trial by the Inquisition. & Landini, Firenze, 1632; Wikisource transcription. \\[2pt]
\emph{Confessiones} (Book XIII) & Augustine of Hippo & c.~400 & LAT & 38\,/\,22 & paragraph / chapter & Final book of the spiritual autobiography: a meditation on the opening verses of Genesis and on grace in self-knowledge. & J.-P.\ Migne, \emph{PL} 32; Latin Library. \\[2pt]
\emph{De mortibus persecutorum} & Lactantius & c.~315 & LAT & 52\,/\,29 & chapter / grouped & ``On the deaths of the persecutors.'' Pamphlet narrating the violent ends of Roman emperors who had persecuted Christians, Nero to Diocletian. & Creed (ed.), Oxford 1984; Latin Library. \\[2pt]
\emph{Inferno} & Dante Alighieri & c.~1320 & IT & 34 & canto & First part of the \emph{Divina Commedia}; Dante's descent through 34 cantos of Hell, guided by Virgil. & Petrocchi critical ed.; Wikisource. \\[2pt]
\emph{Heptam\'eron} & Marguerite de Navarre & 1558--59 & FR & 72 & nouvelle & 72 framed nouvelles told by ten travellers stranded in a Pyrenean abbey, each followed by group discussion; French counterpart to the \emph{Decameron}. & Gruget, Paris, 1559; Wikisource. \\[2pt]
\emph{De consolatione philosophiae} & Boethius & c.~524 & LAT & 72 & prose/verse piece & Philosophical dialogue from prison: Lady Philosophy visits the condemned Boethius and reasons through fortune, providence, and the highest good. & Loeb Classical Library; Latin Library. \\[2pt]
\emph{Epigrammata} & Ausonius of Bordeaux & c.~370 & LAT & 27 & individual poem & 27 short occasional epigrams from the late-Latin poet's collected works. & Aldine \emph{Opera}, Venice, 1517; author's OCR. \\
\bottomrule
\end{tabular}
\caption{Primary corpus of natural text-graph instances. Where two $N$ values appear separated by ``$/$'' (Giambullari, Augustine, Lactantius), the row's \emph{Unit} column lists which chunking serves which use.}
\label{tab:corpus}
\end{table}

\noindent The texts span four centuries (4th--17th), three languages (Latin, Italian, French), and deliberately heterogeneous genres. The ``Unit'' column reports the segmentation that determines $N$ following the tiered granularity rule of \S\ref{sec:pipeline}: authorial divisions where they exist, printer's-page fall-back where they do not, and fixed analytic windows for continuous prose (Galileo). Several texts appear at more than one granularity, each table using the one most informative for its argument: \emph{Giambullari} $N{=}151$ for rigidity, $N{=}65$ on the QPU; \emph{Augustine} $N{=}38$ (paragraph) throughout including QPU, $N{=}22$ (chapter) in classical benchmarking only; \emph{Lactantius} $N{=}52$ (chapter) throughout, $N{=}29$ (grouped) in benchmarking only; \emph{Dante} $N{=}34$ (canto) for the \emph{Inferno}, $N{=}100$ for the full \emph{Divina Commedia} in benchmarking.
\end{landscape}

\section{Embedding validation}\label{app:embedder_validation}

\vspace{0.5em}

The paper fixes a single embedder for natural-text graphs --- \texttt{intfloat/\allowbreak multilingual-\allowbreak e5-\allowbreak large-\allowbreak instruct} \cite{wang2024multilinguale5instruct,enevoldsen2025mmteb}, top-ranked on the MMTEB multilingual benchmark and handling Latin, Italian, and French without language-specific tuning. The natural question is whether a different embedder would change the structural conclusions: are $\mathrm{MIS}/N$ and $\rho$ stable across embedders, or are they artefacts of the e5 family?

To answer this we re-ran the full pipeline on four natural texts spanning languages and lengths (Dante's \emph{Inferno}, $N{=}68$, Italian; Boethius' \emph{Consolatio Philosophiae}, $N{=}72$, Latin; Galileo's \emph{Dialogo}, $N{=}130$, Italian; Giambullari, $N{=}151$, Italian --- the cross-embedder check uses a finer chunking than the rest of the paper for \emph{Inferno} (half-canto, $N{=}68$ instead of the canto-level $N{=}34$ used elsewhere) and \emph{Dialogo} (200-word window, $N{=}130$ instead of the 400-word $N{=}65$ of \S\ref{sec:quantum}), so that each text has enough nodes to give the rigidity test statistical resolution under all three embedders) with three competitive multilingual embedders: \texttt{intfloat/\allowbreak multilingual-\allowbreak e5-\allowbreak large-\allowbreak instruct} \cite{wang2024multilinguale5instruct}, \texttt{BAAI/\allowbreak bge-m3} \cite{chen2024bgem3}, and \texttt{Snowflake/\allowbreak snowflake-\allowbreak arctic-\allowbreak embed-\allowbreak l-v2.0} \cite{yu2024arcticembed}. Each embedder feeds the same downstream pipeline --- cosine $k$-NN at $k{=}8$, exact ILP MIS, and per-vertex ILP-exclusion for $\rho$ ($t_\text{lim} = 15$~s per vertex, 32-way parallel) --- so only the embedding model varies. Table~\ref{tab:cross_embedder} gives the side-by-side numbers.

\vspace{0.3em}\noindent\emph{Reading the table.} Table~\ref{tab:cross_embedder} is designed for \emph{within-row} comparison across the three embedders and is run as a self-contained sensitivity experiment with its own graph-build settings (in particular, a symmetric-union $k$-NN graph construction that yields denser edge sets than the canonical mutual-$k$-NN build used in Tables~\ref{tab:dh_summary} and \ref{tab:corpus}). Absolute MIS, $|E|$, and $\rho$ values in Table~\ref{tab:cross_embedder} therefore differ from the canonical Table~\ref{tab:dh_summary} numbers even on the same text at the same $N$ and $k$ (e.g.\ Boethius $N{=}72$, $k{=}8$: $\rho = 0.000$ canonical vs $\rho = 0.333$ here, on a denser graph), and the two tables should not be cross-compared cell-by-cell. The single conclusion this appendix supports is the within-row one: across the three embedders, $\mathrm{MIS}/N$ is stable, $\rho$ is not.

\begin{table}[!h]
\centering
\small
\begin{tabular}{llcccc}
\toprule
\textbf{Text} & \textbf{Embedder} & $|E|$ & \textbf{MIS} & $\mathrm{MIS}/N$ & $\rho$ \\
\midrule
Dante \emph{Inferno} ($N{=}68$)        & e5-large-instruct  & 416 & 21 & 0.309 & 0.476 \\
                                       & bge-m3             & 405 & 17 & 0.250 & 0.118 \\
                                       & arctic-embed-l-v2  & 411 & 18 & 0.265 & 0.444 \\
\midrule
Boethius \emph{Consolatio} ($N{=}72$)  & e5-large-instruct  & 399 & 18 & 0.250 & 0.333 \\
                                       & bge-m3             & 401 & 18 & 0.250 & 0.444 \\
                                       & arctic-embed-l-v2  & 403 & 19 & 0.264 & 0.474 \\
\midrule
Galileo \emph{Dialogo} ($N{=}130$)     & e5-large-instruct  & 829 & 40 & 0.308 & 0.475 \\
                                       & bge-m3             & 815 & 37 & 0.285 & 0.297 \\
                                       & arctic-embed-l-v2  & 765 & 34 & 0.262 & 0.706 \\
\midrule
Giambullari ($N{=}151$)                & e5-large-instruct  & 864 & 40 & 0.265 & 0.325 \\
                                       & bge-m3             & 894 & 43 & 0.285 & 0.558 \\
                                       & arctic-embed-l-v2  & 923 & 41 & 0.272 & 0.195 \\
\bottomrule
\end{tabular}
\caption{Cross-embedder sensitivity on four natural texts. Each text is run through three multilingual transformers; downstream pipeline ($k$-NN at $k{=}8$, ILP MIS, certified $\rho$) is held fixed. $\mathrm{MIS}/N$ stays in the $0.25$--$0.31$ band across all twelve (text, embedder) pairs; $\rho$ varies from $0.118$ to $0.706$ and reorders the texts.}
\label{tab:cross_embedder}
\end{table}

Two findings emerge. First, $\mathrm{MIS}/N$ is embedder-stable: the spread within each text is at most $0.06$, and all four texts cluster in the same $0.25$--$0.31$ band irrespective of which transformer produced the embedding. Backbone \emph{size} is therefore a robust property of the text. Second, $\rho$ is not stable, and the instability is informative rather than noisy: the embedders disagree on which text is the most rigid (Galileo at $\rho=0.71$ under \texttt{arctic-embed-l-v2} drops to $\rho=0.30$ under \texttt{bge-m3}), and the disagreement preserves the qualitative spectrum --- semi-rigid versus modular --- while permuting the ranking. The interpretation is that $\rho$ is a property of the \emph{(text, embedder)} pair, not of the text alone, and any cross-corpus comparison of rigidity should fix the embedder. The paper fixes \texttt{e5-large-instruct} throughout for consistency, motivated by its top rank on the MMTEB multilingual benchmark \cite{enevoldsen2025mmteb}.

\section{\emph{Twelve Hours of a Coastal Town}: full text}\label{app:twelve_hours}

Complete text of the twelve-paragraph mini-text used for the worked example of \S\ref{subsec:worked_example}. The MIS backbone is $\{$¶3, ¶4, ¶7, ¶10$\}$, one paragraph from each of the four thematic clusters (dawn departure, mid-morning market, silent noon, late-afternoon return); the persistent core is $\{$¶3, ¶10$\}$ --- the cliff-watcher and the boy mending nets, present in every one of the nine optimal backbones.

\begin{quote}\small
\textbf{(¶1)} \emph{At first light the fishermen drag their wooden boats across the wet sand. The hulls leave parallel grooves that the next tide will erase. Nets are loaded, oars are checked, the youngest boy carries the bait bucket without complaint.}\\[2pt]
\textbf{(¶2)} \emph{The harbour wakes by degrees. A second crew of fishermen pushes off from the stone jetty, their boat heavier than the first, their nets longer. Gulls follow them out, screaming for what has already been thrown overboard.}\\[2pt]
\textbf{(¶3)} \emph{From the cliff above the beach an old fisherman watches the two crews vanish. He no longer goes to sea. He keeps a small notebook of weather signs and writes one line each morning, always about wind direction and the colour of the horizon.}\\[2pt]
\textbf{(¶4)} \emph{By mid-morning the market square is loud with bargaining. Fishwives shout the prices of mackerel and sardines, butchers slap the boards with their cleavers, the children of the town run between the stalls stealing peaches when no one is looking.}\\[2pt]
\textbf{(¶5)} \emph{At the centre of the market a cheese-seller has set up a long table covered with white cloths. She arranges the rounds of pecorino in pyramids and offers thin slices on the point of a knife to anyone who looks hungry. The bargaining around her stall is louder than anywhere else.}\\[2pt]
\textbf{(¶6)} \emph{The market smells of fish and lemon and warm bread. Two priests pass through on their way to the harbour chapel, refusing to bargain for anything, blessing the children who try to steal their hats. The square will not empty until the heat of midday makes everyone retreat indoors.}\\[2pt]
\textbf{(¶7)} \emph{After lunch the town falls silent. The shutters close one by one, the dogs sleep in the shadow of the church wall, and the only sound is the rope of a small boat creaking against its bollard down at the harbour. Even the gulls have stopped screaming.}\\[2pt]
\textbf{(¶8)} \emph{A single cat walks the empty length of the main street, stops, looks at nothing in particular, then disappears under a parked cart. The shadow of the bell tower has reached the fountain. No one draws water; the noon hour belongs to no one.}\\[2pt]
\textbf{(¶9)} \emph{In the orchard behind the church a beekeeper turns his hives one quarter turn east, as he does every Tuesday afternoon while the town sleeps. The bees do not protest. He works in silence, bare-handed, and is back inside his cool kitchen before the first shutter reopens.}\\[2pt]
\textbf{(¶10)} \emph{By late afternoon the tide has come in and is licking the keels of the boats hauled up that morning. The same boy who carried the bait bucket at dawn now sits on the gunwale of his father's boat, mending a torn net with a wooden needle, his fingers already as quick as a man's.}\\[2pt]
\textbf{(¶11)} \emph{The second crew returns rowing slowly, their boat low in the water with the day's catch. The harbour wall is lined with women and children waiting to count the baskets. The old fisherman from the cliff has come down at last, and stands at the back of the crowd, saying nothing.}\\[2pt]
\textbf{(¶12)} \emph{At sunset the harbour is loud again, but it is a different loudness from the morning. Lanterns are lit on the jetty, the fish are sorted into baskets by size and not by species, and the cheese-seller from the market square walks down with two empty crates to buy a share of the catch for tomorrow.}
\end{quote}

\section{Classical-benchmark results}\label{app:bench_results}

\vspace{0.5em}

We benchmarked three classical MIS solvers on all text-graph instances --- exact integer linear programming (ILP), a greedy heuristic (maximum-degree-first), and simulated annealing --- and report the per-instance numbers in Table~\ref{tab:classical_benchmark}.

The hardest single-instance ILP solve in the natural-text corpus is \emph{Boethius} ($k{=}8$, $N{=}72$, ${\geq}500$ distinct optima): the MIS itself is found in $0.06$~s, but exhaustive enumeration of the optimal-MIS manifold up to a $500$-cap takes ${\sim}\,$$106$~s of cumulative ILP time ($^{\dag\dag}$). \emph{Heptam\'eron} at $k{=}16$ is the single-instance $1.5$~s case ($^{\dag}$, MIS plus 6 distinct optima). The hardest QOuLiPo engineered instance is \emph{Le Pari de Nithard III} ($k{=}16$, $N{=}100$); the small-$N$ standout is \emph{Sonetti dal Tesseratto} ($N{=}16$, $d{=}0.658$, the densest QOuLiPo graph, where the $Q_4$ hypercube $k$-NN saturation is genuinely harder than its size suggests). The exact-UDG QOuLiPo rows (Stanze, Castello, Treize, Squares, Faces, Hours, Kagome, Postman) are omitted from the table as classically trivial: 2D-UDG MIS admits a PTAS, and CBC solves each in $<0.1$~s. Erd\H{o}s--R\'enyi random graphs at the same density are 20--40$\times$ harder than the engineered texts: the engineered structure regularises the MIS landscape that ILP exploits.

\begin{table}[H]
\centering
\small
\begin{tabular}{lcccccc}
\toprule
\textbf{Graph} & \textbf{$N$} & $d$ & \textbf{ILP (exact)} & $t_\text{ILP}$ & \textbf{Greedy} & \textbf{SA} \\
\midrule
Augustine ($k_\text{nn}{=}8$) & 38 & 0.293 & 9 & 0.12~s & 7 & 9 \\
Lactantius ($k_\text{nn}{=}8$) & 52 & 0.232 & 13 & 0.07~s & 11 & 13 \\
Heptam\'eron ($k_\text{nn}{=}16$) & 72 & 0.331 & 14 & 1.5~s$^{\dag}$ & --- & --- \\
Boethius ($k_\text{nn}{=}8$) & 72 & 0.146 & 14 & 106~s$^{\dag\dag}$ & --- & --- \\
Nithard's Wager ($k{=}16$) & 50 & 0.305 & 14 & 0.07~s & 13 & 14 \\
Le Pari de Nithard ($k{=}16$) & 50 & 0.349 & 10 & 0.08~s & 10 & 10 \\
Le Pari de Nithard~II ($k=16$) & 65 & 0.303 & 8 & 1.0~s & 8 & 7 \\
Galileo 65 ($k=16$) & 65 & 0.249 & 14 & 0.16~s & 13 & 13 \\
Giambullari 65 ($k=8$) & 65 & 0.120 & 20 & 0.06~s & 20 & 19 \\
Giambullari 151 (full) & 151 & 0.077 & 41 & 2.3~s & 40 & 40 \\
\midrule
Le Pari de Nithard~III ($k=16$) & 100 & 0.222 & 20 & 1.5~s & 18 & 17 \\
Le Pari de Nithard~III ($k=25$) & 100 & 0.336 & 15 & 1.0~s & 13 & 12 \\
Sonetti dal Tesseratto ($Q_4$, $k{=}8$) & 16 & 0.658 & 4 & 0.72~s & --- & --- \\
\midrule
\emph{ER random ($d \approx 0.3$)} & 100 & 0.298 & 15 & \textbf{33~s} & 14 & 12 \\
\emph{ER random ($d \approx 0.4$)} & 100 & 0.399 & 11 & \textbf{42~s} & 11 & 10 \\
\emph{ER random ($d \approx 0.5$)} & 100 & 0.495 & 9 & \textbf{46~s} & 7 & 7 \\
\bottomrule
\end{tabular}
\caption{Classical MIS benchmark (CBC ILP). $^{\dag}$ \emph{Heptam\'eron}: time includes MIS plus enumeration of $6$ distinct optima. $^{\dag\dag}$ \emph{Boethius}: MIS in $0.06$~s; the reported $106$~s is cumulative ILP time for enumerating the optimal-MIS manifold up to the $500$-solution cap.}
\label{tab:classical_benchmark}
\end{table}

\clearpage
\begin{landscape}
\section{QOuLiPo: full corpus inventory}\label{app:oulipo_full_table}

\vspace{0.5em}

Table~\ref{tab:oulipo_texts} catalogues the twenty-nine engineered QOuLiPo texts, sorted by $N$. Rigidity $\rho$ is the fraction of MIS pages common to every optimal solution (entries showing $\geq$ hit the 500-solution enumeration cap). Source data: \cite{jurczak2026qoulipo_dataset}.

\begin{table}[H]
\centering
\scriptsize
\setlength{\tabcolsep}{6pt}
\begin{tabular}{lcrrrrrrrccl@{}}
\toprule
\textbf{Text} & \textbf{Lang} & \textbf{$N$} & \textbf{$|E|$} & \textbf{$d$} & \textbf{$\rho$} & \textbf{MIS} & \textbf{MIS/$N$} & \textbf{\#\,opt.} & \textbf{UDG} & \textbf{HW} & \textbf{Designed property} \\
\midrule
Sonetti dal Tesseratto              & IT    &  16 &  79 & 0.658 & 1.000 &  4 & 0.250 &  1            &              & EMU   & $Q_4$ hypercube; 16 sonnets at $k{=}8$ \\
Les Aventures de Pascal             & FR    &  20 &  47 & 0.247 & 0.000 &  5 & 0.250 &  4            & \checkmark   & EMU\,(2L)   & King $5{\times}4$, single layer of bilayer pair \\
Scholia de M\'enil                  & FR    &  20 &  47 & 0.247 & 0.000 &  5 & 0.250 &  5            & \checkmark   & EMU\,(2L)   & King $5{\times}4$, single layer (FR commentary) \\
The Friar's Notebook                & EN    &  20 &  30 & 0.158 & 0.000 &  8 & 0.400 &  5            &              & EMU\,(3D)   & Pacioli dodecahedron, 30 register-aware motifs (3D UBG) \\
La Vita nel Cubo                    & IT    &  23 & 128 & 0.506 & 0.571 &  7 & 0.304 &  5            &              & EMU   & Truncated-cube conceit, $k{=}8$ \\
Le venticinque stanze               & IT    &  25 &  72 & 0.240 & 1.000 &  9 & 0.360 &  1            & \checkmark   & QPU   & King $5{\times}5$ \\
The Twenty-Five Rooms               & EN    &  25 &  72 & 0.240 & 1.000 &  9 & 0.360 &  1            & \checkmark   & EMU   & King $5{\times}5$ (EN twin of \emph{venticinque stanze}) \\
Les Jumeaux du Graphe (EN)          & EN    &  30 &  38 & 0.087 & 0.000 & 14 & 0.467 & 21            &              & EMU   & $C_{30}$ + 8 same-parity chords \\
Les Jumeaux du Graphe (FR)          & FR    &  30 &  38 & 0.087 & 0.000 & 14 & 0.467 & 21            &              & EMU   & FR twin of EN, shared edge list \\
Treize sur trente-sept              & FR    &  37 &  90 & 0.135 & 1.000 & 13 & 0.351 &  1            & \checkmark   & QPU   & Centered-hex patch, $R{=}3$ \\
Pascal--M\'enil bilayer             & FR    &  40 & 240 & 0.308 & 0.000 &  6 & 0.150 & $\geq 500$    &              & EMU\,(2L)   & $5{\times}4{\times}2$ king bilayer (UBG) \\
La Partition du Texte               & FR    &  46 & 244 & 0.236 & 0.091 & 11 & 0.239 & 25            &              & EMU   & Five poetic-form communities, $k{=}8$ \\
Il castello dei quarantanove destini& IT    &  49 & 396 & 0.337 & 1.000 &  9 & 0.184 &  1            & \checkmark   & QPU   & Extended-king $7{\times}7$ (Cheb. ${\leq}2$); king-only on QPU \\
Nithard's Wager (EN)                & EN    &  50 & 374 & 0.305 & 0.571 & 14 & 0.280 &  6            &              & QPU   & Hard zone, $k{=}16$ \\
Le Pari de Nithard (FR translation) & FR    &  50 & 427 & 0.349 & 0.400 & 10 & 0.200 & 19            &              & QPU   & Hard zone, $k{=}16$ \\
The Irreplaceable Book              & EN    &  50 & 290 & 0.237 & 1.000 & 17 & 0.340 &  1            &              & EMU   & Unique MIS via double domination; max degree 19 (not 2D-UDG-realisable) \\
The Map of the Text                 & EN    &  50 &  85 & 0.069 & 0.000 & 25 & 0.500 &  2            & \checkmark   & EMU   & Planar grid $5{\times}10$ \\
The Trial of Nithard                & EN    &  50 & 625 & 0.510 & 0.000 & 25 & 0.500 &  2            &              & EMU   & Bipartite $K_{25,25}$ (provably non-UDG) \\
The Reader's Trap                   & EN    &  50 &  72 & 0.059 & 0.417 & 24 & 0.480 & $\geq 500$    & \checkmark   & QPU   & Hub--spoke; greedy-vs-optimum gap maximised \\
The Fractal Book                    & EN    &  50 & 102 & 0.083 & 0.316 & 19 & 0.380 & $\geq 267$    & \checkmark   & QPU   & Self-similar (Sierpinski) \\
The Kaleidoscope                    & EN    &  51 &  51 & 0.040 & 0.000 & 17 & 0.333 & $3^{17}$      & \checkmark   & EMU   & Maximal degeneracy: 17 disjoint $K_3$ \\
The Incarnate Graph                 & EN    &  65 & 155 & 0.075 & 0.000 & 22 & 0.338 & $\geq 385$    & \checkmark   & QPU   & Random UDG placement (insufficient blockade margin) \\
Le Pari de Nithard~II               & FR    &  65 & 631 & 0.303 & 0.250 &  8 & 0.123 & $\geq 255$    &              & QPU   & Hard zone at $N{=}65$, $k{=}16$ \\
The Book of Eighty-One Squares      & EN    &  81 & 272 & 0.084 & 1.000 & 25 & 0.309 &  1            & \checkmark   & QPU   & King $9{\times}9$ \\
Eighty-One Faces of a Pomegranate   & EN    &  81 & 622 & 0.192 & 1.000 & 13 & 0.160 &  1            & \checkmark   & QPU   & Extended-king $9{\times}9$ ($\sqrt{5}$ reach) \\
Hours of the Day                    & EN    &  91 & 240 & 0.059 & 1.000 & 31 & 0.341 &  1            & \checkmark   & QPU   & Centered-hex patch, $R{=}5$ \\
Kagome of Late Summer               & EN    & 100 & 180 & 0.036 & 0.846 & 39 & 0.390 &  8            & \checkmark   & QPU   & Kagome (3.6.3.6) Archimedean tiling \\
The Postman's Songbook              & EN    & 100 & 220 & 0.044 & 0.333 & 36 & 0.360 & 97            & \checkmark   & QPU   & Snub-square (3.3.4.3.4) Archimedean tiling \\
Le Pari de Nithard~III              & FR    & 100 & 691 & 0.140 & 0.923 & 26 & 0.260 &  4            &              & EMU             & 100-atom designed $k$-NN target, $k{=}8$ \\
\bottomrule
\end{tabular}
\caption{Full QOuLiPo corpus inventory (twenty-nine constrained texts), sorted by register size $N$. \textsc{HW} codes: \texttt{QPU} = adiabatic FRESNEL\_CAN1 run; \texttt{EMU} = EMU\_MPS sweep (Pasqal Cloud A100 back-end for most runs, commercial H100/H200 cloud for the larger instances); \texttt{EMU\,(2L)} and \texttt{EMU\,(3D)} = bilayer / 3D EMU runs.}
\label{tab:oulipo_texts}
\end{table}

\end{landscape}

% ─── Appendix F: Hardware-side technical detail ─────────────────────────

\section{Register and pulse optimisation}\label{app:hardware_detail}

This appendix gives the detailed accounting of register engineering and pulse design summarised in \S\ref{subsec:fresnel_setup}.

\subsection*{F.1 Register engineering}

The register embedding proceeds differently for the two halves of the corpus. For natural texts $G_{\text{text}}$ is not unit-disk in general, so it must be mapped onto physical atom positions --- a non-trivial geometric optimisation that produces $G_{\text{reg}}$ as the lossy realisation. We use simulated annealing (SA) on discrete sites of a triangular lattice, which handles the non-convex landscape of blockade constraints robustly without requiring gradient information, and which Dalyac and Henriet \cite{dalyac2022embed3d} also adopt for neutral-atom register embedding. The SA embedder places $N$ atoms on the lattice such that atom pairs within the blockade radius $R_b = 8.0$~$\mu$m ($G_{\text{reg}}$ edges) correspond as closely as possible to $G_{\text{text}}$ edges; the optimiser runs 16{,}000--20{,}000 iterations with 6--10 random restarts (move + swap moves), maximising \emph{edge recall} --- the fraction of canonical edges $(u, v) \in G_{\text{text}}$ whose embedded coordinates lie within $R_b$ in $G_{\text{reg}}$, the standard information-retrieval recall metric \cite{manning2008ir} applied to graph adjacency. For the engineered exact-UDG texts the register embedding is trivially exact: place atoms at integer grid coordinates with spacing $d$ and set $R_b = d\sqrt{2} + \epsilon$. Every king adjacency falls within $R_b$; every non-adjacent pair is at distance $\geq 2d$, outside $R_b$. Edge recall is $1.0$ by construction ($G_{\text{design}} = G_{\text{reg}}$), no SA needed.

\begin{figure}[!htbp]
\centering
\includegraphics[width=\textwidth]{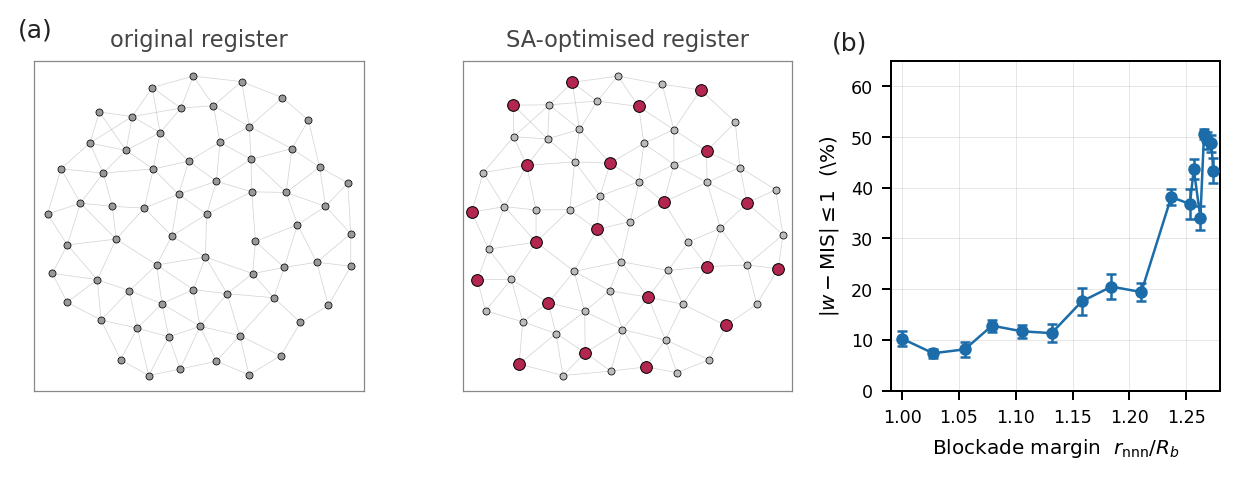}
\caption{Register engineering on \emph{The Incarnate Graph} ($N=65$, random UDG). (a) Two QPU runs on the same canonical graph and the same baseline pulse, two different registers. Left: original 2D placement, $0$ valid IS out of $1000$ shots (ratio $0.000$). Right: SA-optimised placement, with the largest valid independent set returned by the QPU shown in red ($21/22$, ratio $0.955$, $402$ shots). (b) Blockade-margin curve on the same canonical graph: $|w-\mathrm{MIS}|\leq 1$ fraction --- the fraction of shots whose Hamming weight is within one of the exact MIS size --- versus achieved blockade margin $r_{\text{nnn}}/R_b$, EMU simulation.}
\label{fig:register_engineering}
\end{figure}

Edge recall $1.0$ by construction is necessary but not sufficient for clean MIS recovery on hardware, as \emph{The Incarnate Graph} ($N{=}65$, random-placement exact 2D UDG) makes plain: $0/1000$ valid IS (i.e.\ shots whose bitstring forms an independent set on the canonical graph), with the Hamming-weight distribution centred near $\mathrm{HW} \approx 18$ (close to $\mathrm{MIS}{=}22$) --- the system reaches the right occupation count but cannot satisfy independence. The reason is geometric: an exact-UDG canonical graph is necessary but not sufficient for clean MIS recovery, because the register --- the physical placement of atoms --- must also satisfy a sufficient \emph{blockade margin}, which we use as shorthand for the ratio $r_{\text{nnn}}/R_b$ between the smallest non-edge (next-nearest-neighbour) distance and the blockade radius. The qualitative requirement $r_{\text{nn}} \ll R_b \ll r_{\text{nnn}}$ for clean Rydberg-blockade embeddings of unit-disk graphs is well-established (Dalyac \emph{et al.}\ \cite{dalyac2024graph}, Box~1). The practical implication is that when a register layout fails on hardware (zero valid IS, as on the original \emph{Incarnate} run --- see the case studies below), one can often recover it by repositioning the atoms --- keeping the canonical graph and the pulse fixed --- so that all non-edges sit comfortably above $R_b$. How far above is achievable is graph-dependent: each canonical graph has its own geometric ceiling on $r_{\text{nnn}}/R_b$, beyond which no register can be found.

\emph{Treize sur trente-sept} (centred-hex axial-coordinate bug, refit to a proper triangular layout) recovers to ratio $1.000$ on EMU and $0.923$ on QPU; \emph{The Kaleidoscope} (17 disjoint $K_3$, refit to hex-packed cluster centres) recovers to $1.000$ on EMU. On \emph{The Incarnate Graph} (Figure~\ref{fig:register_engineering}a) the original random UDG placement left $17$ non-edges in the soft tail at $\leq 1.1\,R_b$; an SA register pushing all non-edges beyond $1.25\,R_b$ rescues the row on QPU. Re-spacing the \emph{Hours of the Day} lattice ($N{=}91$ centred-hex) from $a{=}5\,\mu$m to $a{=}6\,\mu$m similarly lifts the blockade margin from $1.08$ to $1.30$, taking the row from $0/1000$ valid IS to ratio $0.839$ at $23.9\%$ valid IS on $1000$ shots. The same approach was used to design two non-lattice 2D-UDG instances directly for the QPU: \emph{The Reader's Trap} ($1.000 / 65.4\%$) and \emph{The Fractal Book} ($0.947 / 60.5\%$).

Figure~\ref{fig:register_engineering}b quantifies the rule. We swept the achieved blockade margin $r_{\text{nnn}}/R_b$ on the \emph{Incarnate} canonical graph from $1.00$ to $1.27$, re-engineering the register at each target by SA, reusing the same canonical edges and baseline pulse. The $|w-\mathrm{MIS}|\leq 1$ fraction --- a softer cross-platform metric than strict valid IS, well-defined whether the canonical graph is exact-UDG or $k$-NN --- transitions monotonically with the achieved margin, identifies an empirically sufficient threshold for clean recovery on this graph, and exposes the graph's geometric saturation cap beyond which no register can do better.

Register engineering, especially for non-lattice canonical graphs, is a necessary step rather than a cosmetic one. SA on atom positions suffices for every case here; further techniques are surveyed in \cite{dalyac2024graph}.

\subsection*{F.2 Pulse optimisation}

The atoms are driven by a simple baseline pulse, constructed with Pasqal's open-source Pulser library \cite{silverio2022pulser}: constant Rabi frequency $\Omega = C_6 / R_b^6 = 3.30$~rad/$\mu$s and a linear detuning sweep from $-3\Omega$ to $+3\Omega$ over $T = 4~\mu$s, chosen for reproducibility rather than optimality. For the larger instances ($N \geq 81$) we extend the ramp to $T = 6\,\mu$s --- the QPU's adiabatic-pulse ceiling --- because the inverse spectral gap of the MIS scales with system size, and the longer ramp is necessary for the system to follow the instantaneous ground state through the gap-closing region into the MIS configuration.

The pulse-engineering axis is in principle fertile, and the EMU loop of Figure~\ref{fig:qpu_pipeline} is well-suited to exploring it cheaply. We ran a 10-shape variational sweep --- baseline, longer $T$, wider/tighter detuning sweep, $\Omega$-trapezoid envelope, four-knot asymmetric ramps in the spirit of the variational adiabatic shaping of Ebadi \emph{et al.}\ \cite{ebadi2022quantum}, multi-point cubic detuning interpolation, and jittered-register controls --- on EMU for the strong-tier exact-UDG rows. On EMU the variational shapes give substantial lifts on the validity and structural-readout indicators while leaving the ratio at $1.000$. On QPU the corresponding lifts are more modest: EMU-selected variational pulses improve \texttt{valid IS\%} by $+3$--$6$ percentage points over baseline, ratio pinned at $1.000$. \emph{The Reader's Trap} and \emph{The Fractal Book} (the two non-lattice 2D-UDG instances) ran with an $\Omega$ rise-plateau-fall envelope rather than the flat baseline and cleared $60\%$ valid IS at ratio $\geq 0.95$. For \emph{The Postman's Songbook} (snub-square Archimedean tiling) the multi-distance geometry required reducing $\Omega$ to $1.63$~rad/$\mu$s and proportionally rescaling the detuning span; the standard $\Omega = 3.30$ pulse over-drives the next-nearest pairs of the snub-square lattice. On natural-text rows the trade-off is more nuanced: pulse variants that improve near-valid coverage tend to reduce the strict approximation ratio, and vice versa. Pulse engineering at FRESNEL scale is therefore a second-order refinement on top of a working register, not a substitute for one. Noise-aware pulse optimisation \cite{cazals2025mis} and Floquet / dynamical-decoupling pulse design \cite{dalyac2024graph} are natural follow-ups.

\end{document}